\documentclass[preprint,amsfonts,amsmath,pra,aps,letterpaper]{revtex4}
\usepackage[T1]{fontenc}
\usepackage[dvips]{epsfig}
\usepackage[dvips]{graphicx}
\usepackage{psfrag}
\usepackage{rotating}
\usepackage{bm}       
\usepackage{amssymb}  
\usepackage{mathrsfs}
\usepackage{bm}
\usepackage{bbold}
\usepackage{braket} 

\newcommand{\gvec}[1]{\ensuremath{\mathbf{#1}}}
\newcommand{\uvec}[1]{\ensuremath{\hat{\gvec{#1}}}}

\begin{document}
\date{\today}
\flushbottom 
\title{
Collisional decoherence of internal state superpositions in a trapped ultracold gas
}
\author{C. J. Hemming and R. V. Krems}
\affiliation{
Department of Chemistry, University of British Columbia, Vancouver, B.C. V6T 1Z1, Canada
}
\begin{abstract}
We analyze collisional decoherence of atoms or molecules prepared in a coherent superposition of nondegenerate internal states at ultralow temperatures and placed in an ultracold buffer gas.  Our analysis is applicable for an arbitrary bath particle/tracer particle mass ratio. Both elastic and inelastic collisions contribute to decoherence.  We obtain an expression relating the observable decoherence rate to pairwise scattering properties, specifically the low-temperature scattering amplitudes.  We consider the dependence on the bath particle/tracer particle mass ratio for the case of light bath and heavy tracer particles.  The expressions obtained may be useful in low-temperature applications where accurate estimates of decoherence rates are needed.  The results suggest a method for determining the scattering lengths of atoms and molecules in different internal states by measuring decoherence-induced damping of coherent oscillations.
\end{abstract}

\maketitle

\section{Introduction}

Atomic and molecular interferometry experiments \cite{inter1,inter2}, precision measurements of fundamental constants \cite{precision1,precision2,precision3,demille:pra} and coherent control of molecular dynamics \cite{cc1} are based on gaseous ensembles of atoms or molecules prepared in coherent superpositions of internal (electronic, ro-vibrational, hyperfine, Stark or Zeeman) energy states. Atoms and molecules in coherent superpositions of internal states have also been proposed as building blocks for quantum computation and quantum information processing \cite{qc1,qc2}. Coherent superpositions may be destroyed by external field fluctuations and collisions of gas particles. Collision-induced decoherence is a major limiting factor in the experimental realization of quantum computation and coherent control of molecular dynamics. Recent progress in the development of experimental techniques for cooling atoms and molecules to extremely low temperatures suggests new possibilities for precision spectroscopy measurements, coherent control of molecular processes and quantum computation \cite{review}. For example, cooling molecules to ultralow temperatures allows for high-resolution spectroscopy with long interrogation times and high degree of control over intra- and inter-molecular interactions \cite{precision4}. The translational energy of ultracold molecules is insignificant and can be disentangled from internal states, which can be exploited to develop new schemes for coherent control \cite{felipe}. However, elastic and inelastic collisions of atoms and molecules at ultralow temperatures may be very efficient, leading to  significant decoherence rates \cite{book}. In order to assess the feasibility of quantum information processing, quantum interferometry measurements and coherent control schemes based on ultracold atom and molecules, it is  necessary to develop a microsopic theory of collisional decoherence of internal state superpositions at ultralow temperatures. 
 
A master equation describing decoherence of translational and internal states for a tracer molecule in an inert gas has been derived heuristically by Hornberger and Vacchini \cite{hv:pra}, and the Brownian motion limit of an infinitely massive tracer particle was considered by Hornberger \cite{hornberger:epl}.  This latter result was applied to decoherence of enantiomeric states of optically active molecules to explain Hund's paradox, i.e.\ the well-known observation that chiral molecules are not naturally found in their achiral ground state \cite{trost:prl}.  Although coherence in the tracer molecule's state in general may involve both translational and internal degrees of freedom, in many experiments a substantial simplification arises because the translational degrees of freedom are fully decohered.  Internal state decoherence in this simplified case was studied by Vacchini \cite{vacchini:pra}.  Whether or not the translational degrees of freedom are fully decohered, coherences between internal states typically depend on the translational degrees of freedom.  For an experiment in which only the internal state is probed, the appropriate statistical operator is a reduced density matrix acting on the internal state space, which is obtained by tracing over the translational degrees of freedom.  As a consequence, the evolution of the internal state is typically non-Markovian and the time dependence of the internal state coherence is generally non-exponential \cite{vacchini:pra}.   

The theory of collisional decoherence for a tracer particle with internal states is an extension of the analysis of positional decoherence for a tracer particle without internal states. This problem has been studied in detail \cite{gallis:pra,dodd:prd,diosi:epl,hs,h,hv:pra,vh:physrep}, particularly in the Brownian motion limit of an infinitely massive tracer particle.  Hornberger and Sipe presented a solution for the Brownian motion limit using a convex decomposition of the bath gas density matrix into localized wavepackets to avoid mathematical problems arising because momentum eigenvectors are not normalizable \cite{hs}.  This rigorous method can be used to develop formal replacement rules for consistently handling the singular quantities in the momentum basis.  The replacement rule method was extended to the case of a tracer particle with finite mass \cite{h}.  Adler connected the replacement rules to the method used to resolve the squared Dirac delta function appearing in the Golden Rule derivation \cite{adler}. 

Other work on collisional decoherence of internal state superpositions has included investigation of decoherence suppression using optical pulses \cite{search:pra} and approaches based on Monte Carlo simulations \cite{minami:pra,reinhold:nimprb}.  An analysis by Reinhold and co-workers predicted that measurements of decoherence-induced damping of quantum beats could be used as a sensitive probe of collision cross-sections \cite{reinhold:nimprb}.  Ramakrishna and Seideman investigated decoherence of rotational wavepackets using an approach applicable for dense media \cite{ramakrishna:prl}. This method treats the translational degrees of freedom phenomenologically.  They also found that damping of coherent oscillations and relaxation in the molecular alignment may provide information about elastic and inelastic collision properties.

In this paper we consider decoherence of internal state superpositions for a tracer molecule in the presence of an ultracold buffer gas in a trap.  This process can be studied in buffer gas cooling experiments -- an important class of experiments for the creation of ultracold molecules and precision spectroscopy \cite{weinstein:nature,egorov:epjd,egorov:phd} -- by creating molecular wavepackets or in sympathetic cooling experiments using magneto-optical traps \cite{modugno:science,myatt:prl}.  The separation between internal energy levels of molecules is typically large compared to ultracold trap depths \cite{carr:njp}.  As a consequence, inelastic collisions, in which the internal state of the molecule is changed, release enough energy from internal states into translational motion that both the tracer molecule and the buffer gas atoms involved in the collision are ejected from the trap.  The experiments begin with the translational degrees of freedom of the buffer gas and tracer molecule fully decohered and in thermal equilibrium at the buffer gas temperature  \footnote{This is a common experimental situation; quantum condensates are a noteworthy exception and our analysis in this paper is not applicable to such systems.}.  It will be shown that the translational degrees of freedom remain fully decohered throughout the experiment.   At $t=0$ the tracer molecules are prepared in a coherent superposition of internal states.  We obtain expressions for the coherence present between internal states over time during the experiment in the limit of ultracold temperatures and in the limit of a small but finite buffer gas/tracer molecule mass ratio.  These expressions give the temperature dependence of the decoherence rate at ultracold temperatures in terms of the $s$-wave scattering parameters for collisions between the buffer gas particles and the tracer molecules in the different internal states.  

The number of trapped molecules in experiments with ultracold gases is continually decreasing due to inelastic trap loss.  There are different measures of coherence between internal states that may be of interest depending on the experiment.  In quantum computation, for example, the total strength of the coherent signal relative to the size of the initial ensemble may be important.  In contrast, for an experiment which is observing coherent oscillations arising from interference between internal states, the quantity of interest may be coherence within the trapped collection of molecules, and the decreasing size of the trapped population may be irrelevant.  We derive two different measures of internal state coherence appropriate to each of these experimental scenarios and find the temperature dependence of the decoherence rate at ultracold temperatures.   Our analysis is based on the replacement rule approach of Hornberger, Sipe and Adler \cite{hs,adler}.  It is related to the work of Hornberger in Ref.~\cite{vacchini:pra} which provides a general theory of collisional decoherence of internal states with translational degrees of freedom fully decohered.  Our analysis differs from Ref.~\cite{vacchini:pra} in that we incorporate the presence of inelastic trap loss to model experiments in ultracold traps, and we consider a particular limit of low temperatures.  Our results suggest a method for determining scattering lengths of molecules in different states based on low-temperature measurements of the temperature dependence of decoherence-induced damping of coherent oscillations.

\section{Collisional decoherence in the momentum and internal state representation}

We consider an ensemble of atoms or molecules of mass $M$ prepared at time $t=0$ in an internal state characterized by a reduced density matrix with elements $\rho_{\nu\nu'}(0)$ and placed in a bath of ultracold atoms of mass $m$.  The quantum number $\nu$ denotes an internal energy eigenstate with energy $\epsilon_{\nu}$.  The initial internal state may be pure or mixed.  If there is initially coherence present between internal energy eigenstates, decoherence occurs as a result of collisions with buffer gas atoms.  The goal of this work is to calculate the time evolution of the coherence between internal states $\nu$ and $\nu'$ using the measures of coherence defined in Sec.~\ref{sec:measures}.

We make the following assumptions to model the conditions of experiments with ultracold gases:

1) At $t=0$ the bath particles and the tracer particle translational degrees of freedom are in thermal equilibrium at the ultracold temperature $T$.

2) The tracer particle internal states $\ket{\nu}$ are nondegenerate.      

3) The buffer gas is sufficiently dilute that the time between collisions is long compared to the duration of a collision.  This implies that we can neglect collisions involving three or more particles. 

4) The energy difference between the internal energy levels of the tracer particle is large compared to the trap depth.  As a consequence, inelastic collisions transfer enough internal energy to translational motion to eject both the tracer particle and the buffer gas particle involved in the collision from the trap. 

5) The collisions between buffer gas particles keep them in thermal equilibrium.  In addition, the previous assumption implies that there is no heating of the trapped buffer gas due to inelastic conditions.  As a result, we may treat the buffer gas particle in every collision as being drawn from the thermal equilibrium ensemble at temperature $T$ which does not change in time.

6) The trapping potential is ignored: the trapped gas is treated as a uniform gas in free space.  This is a good approximation since trap dimensions are large compared to relevant length scales for collisions and gas properties typically vary slowly in space.

The tracer particle state at time $t\geq0$ is represented by the reduced density operator $\hat{\rho}(t)$ with matrix elements $\rho_{\nu\nu'}(\gvec{P},\gvec{P}';t)=\braket{\gvec{P}\nu|\hat{\rho}(t)|\gvec{P}'\nu'}$ in the momentum and internal state basis.  
The bath ensemble is described by the thermal equilibrium density matrix corresponding to the bath temperature $T$,
\begin{align}
\hat{\rho}^{\mathrm{gas}} & = \frac{(2\pi\hbar)^3}{\Omega}\int d^3\gvec{p}\,\ket{\gvec{p}}\mu(\gvec{p})\bra{\gvec{p}} \,, \label{eq:bathrho}
\end{align}
where 
\begin{align}
\mu(\gvec{p}) & = \frac{e^{-\frac{p^2}{2mk_BT}}}{(2\pi m k_B T)^{3/2}}\,,\label{eq:mudef}
\end{align}
and $\Omega$ is the system box volume.  $\Omega$ is not to be confused with the trap volume.  It is imagined that the free space system is placed in a box of finite volume $\Omega$ and the continuum limit is obtained as $\Omega\to\infty$.  $\Omega$ appears as a necessary normalization factor.

At $t=0$, the tracer particle density operator is
\begin{align}
\hat{\rho}(0) 
& = \sum_{\nu\nu'}\rho_{\nu\nu'}(0)
\ket{\nu}\bra{\nu'} \otimes \frac{(2\pi\hbar)^3}{\Omega}\int d^3P\, \frac{e^{-\frac{P^2}{2Mk_BT}}}{(2\pi M k_B T)^{3/2}}\ket{\gvec{P}}\bra{\gvec{P}} \,, \label{eq:molinitialconds}
\end{align}
where $\rho_{\nu\nu'}(0)$ are the elements of the reduced density operator which describes the initial internal state, hence $\rho_{\nu\nu'}(0)=\rho^{*}_{\nu'\nu}(0)$ and $0\leq \rho_{\nu\nu'}(0)\rho_{\nu'\nu}(0)/\sqrt{\rho_{\nu\nu}(0)\rho_{\nu'\nu'}(0)} \leq 1$.  

The internal state space of the tracer particle is $\mathcal{H}_\nu$.  The state spaces for the relative translational motion of the colliding pair and the motion of the center of mass  are respectively $\mathcal{H}_{\mathrm{rel}}$  and $\mathcal{H}_{\mathrm{CM}}$.

\subsection{Collisions without trap loss}

The Hamiltonian for the two-particle system consisting of a tracer particle and buffer gas particle is
\begin{equation}
H = H_0 + V\,, \label{eq:hamiltonian}
\end{equation}
where 
\begin{align}
H_0 & = H_0^{\mathrm{m}} + H_0^{\mathrm{gas}} \label{eq:freehamiltonian}
\end{align}
is the sum of the Hamiltonians of the free particles.  $H_0^{\mathrm{m}}$ is the Hamiltonian of the tracer molecule, $H_0^{\mathrm{gas}}$ is the Hamiltonian of the buffer gas atom and $V$ represents the interaction between them. 

At times long before and long after the collision the particles are well separated and the system's evolution is governed by the Hamiltonian $H_0$. If the colliding particles are in a pure quantum state, the two particle state 
$\ket{\Psi(t)}\to e^{-iH_0t/\hbar}\ket{\Psi_{\mathrm{in}}}$ as $t\to-\infty$ for some $\ket{\Psi_{\mathrm{in}}}$, and as $t\to\infty$, $\ket{\Psi(t)}\to e^{-iH_0t/\hbar}\ket{\Psi_{\mathrm{out}}}$.    The incoming and outgoing asymptotic trajectories are related by the two-particle scattering operator $S$ according to \cite{taylor:book}
\begin{equation}
\ket{\Psi_{\mathrm{out}}} = S\ket{\Psi_{\mathrm{in}}}\,. 
\end{equation}
The duration of a collision is short compared to the time scales of the experiments. We may thus regard a collision occurring at time $t$ as an effectively instantaneous transition between the asymptotic trajectories $e^{-iH_0t/\hbar}\ket{\Psi_{\mathrm{in}}}\to e^{-iH_0t/\hbar}\ket{\Psi_{\mathrm{out}}} =  e^{-iH_0t/\hbar}S\ket{\Psi_{\mathrm{in}}} = S e^{-iH_0t/\hbar}\ket{\Psi_{\mathrm{in}}}$.  The second equality is valid because the $S$ operator commutes with $H_0$ \cite{taylor:book}.

If the incoming two-particle state is a mixed state described by a density operator $\hat{\rho}^{\mathrm{pair}}$, then the state after the collision is given by the density operator
\begin{equation}
\hat{\rho}^{\mathrm{pair}\prime} = S\hat{\rho}^{\mathrm{pair}}S^\dag\,.
\end{equation} 
The reduced one particle density operator for the tracer particle state after the collision is obtained by tracing over the degrees of freedom of the bath particle,
\begin{equation}
\hat{\rho}'=\mathrm{tr}_{\mathrm{gas}}\left\{S\hat{\rho}^{\mathrm{pair}}S^\dag\right\}\,.
\end{equation}

A collision will entangle the states of the tracer particle with the buffer gas particle. However, we are interested in describing the ensemble of tracer particles only. This is sufficient for the description of most experiments which do not probe entanglement between the tracer particles and the bath.  It is possible that the tracer particle may encounter the same bath particle more than once but the bath particle will have undergone further collisions entangling its state with those of other bath particles. There is no mechanism which can systematically maintain coherence between the tracer particle and the gas particle state between successive encounters of the same particles.  Thus, for each collision we may treat the two-particle state of the colliding pair as the separable state $\hat{\rho}^{\mathrm{pair}}=\hat{\rho}\otimes\hat{\rho}^{\mathrm{gas}}$, and the tracer particle state after the collision is given in terms of the incoming state $\hat{\rho}$ by
\begin{align}   
\hat{\rho}' & = \mathrm{tr}_{\mathrm{gas}}[ S (\hat{\rho}\otimes\hat{\rho}^{\mathrm{gas}}) S^\dag ] \label{eq:hatrhodef}
\end{align}
where $S$ is the two-particle scattering operator which acts on the space $\mathcal{H}_\nu\otimes\mathcal{H}_{\mathrm{rel}}\otimes\mathcal{H}_{\mathrm{CM}}$, and $\mathrm{tr}_{\mathrm{gas}}$ indicates the trace over the buffer gas particle degrees of freedom.  Collisions are classified as either elastic, in which the internal state does not change, or inelastic, in which it does.  In the absence of trap loss, the total number of molecules is conserved.  Hence, probability is conserved and the $S$ operator is unitary.   

The one-particle scattering operator $S_0$ acts on the space $\mathcal{H}_\nu\otimes\mathcal{H}_{\mathrm{rel}}$ and \cite{taylor:book}
\begin{align}
S & = \mathbb{1}_{\mathrm{CM}}\otimes S_0\,.
\end{align}

The two-particle operator $T$ is related to $S$ by
\begin{align}
S & = \mathbb{1}+iT\,.
\end{align}
Similarly,
\begin{align}
S_0 & = \mathbb{1}+iT_0
\end{align}
where $T=\mathbb{1}_{\mathrm{CM}}\otimes T_0$ and $T_0$ is a one-particle operator with matrix elements
\begin{equation}
\braket{ \mathbf{p}'\nu' | T_0 | \mathbf{p}\nu }  = \frac{m_*}{2\pi\hbar}\delta(E'-E)f_{\nu'\nu}(\gvec{p}',\gvec{p})
\end{equation}
 where $f_{\nu'\nu}(\gvec{p}',\gvec{p})$ is the scattering amplitude for the transition $(\gvec{p}',\nu')\leftarrow(\gvec{p},\nu)$ and $m_*=mM/(m+M)$ is the reduced mass of the colliding pair.  $E'=p^{\prime 2}/(2m_*)+\epsilon_{\nu'}$ and $E=p^2/(2m_*)+\epsilon_\nu$ are the total energies for the states $(\gvec{p}',\nu')$ and $(\gvec{p},\nu)$ respectively.  

The time evolution of $\hat{\rho}$ for tracer particles in free space undergoing collisional decoherence with buffer gas atoms is
\begin{align}
\frac{d}{dt}\hat{\rho} & = -\frac{i}{\hbar}[H_0^{\mathrm{m}},\hat{\rho}] +\left(\frac{d\hat{\rho}}{dt}\right)_{\mathrm{coll}} =
 -\frac{i}{\hbar}[H_0^{\mathrm{m}}+H_{\mathrm{n}},\hat{\rho}]+\mathcal{L}\hat{\rho} \label{eq:generaltimeevol}
\end{align}
where $(d\hat{\rho}/dt)_{\mathrm{coll}}$ arises from collisions
and $\mathcal{L}$ is a dissipative Lindblad operator \cite{h,vacchini:pra,hv:pra,vh:physrep}.  For a tracer particle with internal states, Vacchini and Hornberger present a heuristic derivation of Eq.~\eqref{eq:generaltimeevol} finding \cite{vh:physrep}
\begin{align}
H_{\mathrm{n}} & = -2\pi\hbar^2\frac{n_{\mathrm{gas}}}{m_*}\sum_{\substack{\nu\nu'\\\epsilon_\nu=\epsilon_{\nu'}}}\int d^3p\,\mu(\gvec{p}) \mathrm{Re}\left[f_{\nu\nu'}\left(\mathrm{rel}(\gvec{p},\hat{\gvec{P}}), \mathrm{rel}(\gvec{p},\hat{\gvec{P}})\right)\right]\otimes \ket{\nu}\bra{\nu'} \label{eq:Hn}
\end{align}
and  
\begin{align}
\mathcal{L}\hat{\rho} & =  \sum_{\Delta\epsilon}\int d^3Q\int_{\gvec{Q}\perp}d^2k\,\left[e^{i\gvec{Q}\cdot\hat{\gvec{R}}/\hbar}L(\gvec{k},\hat{\gvec{P}};\gvec{Q},\Delta\epsilon)\hat{\rho}L^\dag(\gvec{k},\hat{\gvec{P}};\gvec{Q},\Delta\epsilon)e^{-i\gvec{Q}\cdot\hat{\gvec{R}}/\hbar}\right]\notag \\
& \qquad -\frac{1}{2}\left[\hat{\rho}L^\dag(\gvec{k},\hat{\gvec{P}};\gvec{Q},\Delta\epsilon)L(\gvec{k},\hat{\gvec{P}};\gvec{Q},\Delta\epsilon) + L^\dag(\gvec{k},\hat{\gvec{P}};\gvec{Q},\Delta\epsilon)L^\dag(\gvec{k},\hat{\gvec{P}};\gvec{Q},\Delta\epsilon)\hat{\rho}\right] \label{eq:lindblad}
\end{align}
where
\begin{multline}
L(\gvec{p},\gvec{P};\gvec{Q},\Delta\epsilon) \\
 = 
\sum_{\substack{\nu\nu'\\\Delta\epsilon=\epsilon_\nu-\epsilon_{\nu'}}} f_{\nu\nu'}\left(\mathrm{rel}(\gvec{p}_{\perp\gvec{Q}},\gvec{P}_{\perp{Q}})-\frac{\gvec{Q}}{2}+\frac{\Delta\epsilon}{Q^2/m_*}\gvec{Q}, 
\mathrm{rel}(\gvec{p}_{\perp\gvec{Q}},\gvec{P}_{\perp{Q}})+\frac{\gvec{Q}}{2}+\frac{\Delta\epsilon}{Q^2/m_*}\gvec{Q} \right) 
\\ 
\times\left[\frac{n_{\mathrm{gas}}m}{m_*^2\gvec{Q}}\mu\left(\gvec{p}_{\perp\gvec{Q}}+\frac{m}{m_*}\frac{\gvec{Q}}{2}+\frac{m}{M}\gvec{P}_{\parallel\gvec{Q}}+\frac{\Delta\epsilon}{Q^2/m_*}\gvec{Q} \right)\right]^{1/2}
\otimes \ket{\nu}\bra{\nu'}
\,.\label{eq:Loperator}
\end{multline}
In Eqs.~\eqref{eq:Hn} to \eqref{eq:Loperator}, $\hat{\gvec{P}}$ and $\hat{\gvec{R}}$ are respectively the momentum and position operators for the tracer molecule,
\begin{equation}
\mathrm{rel}(\mathbf{p},\mathbf{P}) = \frac{m_*}{m}\mathbf{p}-\frac{m_*}{M}\mathbf{P}
\end{equation}
is the relative momentum between a buffer gas atom and a tracer molecule with respective momenta $\gvec{p}$ and $\gvec{P}$, and $\gvec{P}_{\parallel\gvec{Q}}$ and $\gvec{P}_{\perp\gvec{Q}}$ represent the components of $\gvec{P}$ respectively parallel to and  perpendicular to $\gvec{Q}$ (with similar notation for $\gvec{p}_{\perp\gvec{Q}}$).  The integration with respect to $\gvec{k}$ in Eq.~\eqref{eq:lindblad} is over the two-dimensional space perpendicular to $\gvec{Q}$.  

\subsection{Trap loss in inelastic collisions}

We now include trap loss due to inelastic collisions.  The $S$ operator may be decomposed into parts corresponding to elastic and inelastic collisions, as may $\hat{\rho}'$ as given in Eq.~\eqref{eq:hatrhodef}.  We define the elastic part of the $S$ operator 
\begin{align}
S^{\mathrm{el}} & = \sum_{\nu} S_{\nu\nu}
\end{align}
where $S_{\nu\nu'}=\braket{\nu|S|\nu'}$ is an operator on $\mathcal{H}_{\mathrm{rel}}\otimes\mathcal{H}_{\mathrm{CM}}$.  The inelastic part is
\begin{align}
S^{\mathrm{in}} & = S-S^{\mathrm{el}} = \sum_{\nu\neq\nu'}S_{\nu\nu'}\,.
\end{align}
Unlike $S$, the operator $S^{\mathrm{el}}$ is not unitary. 

The $\nu\nu'$ element of $\hat{\rho}(t)$ is $\hat{\rho}_{\nu\nu'}(t)=\braket{\nu|\hat{\rho}(t)|\nu'}$, which is an operator on the tracer particle translational state space.  The elastic term in $\hat{\rho}'$ is
\begin{align}
\hat{\rho}^{\prime\mathrm{el}} & = 
\mathrm{tr}_{\mathrm{gas}}[ S^{\mathrm{el}} (\hat{\rho}\otimes\hat{\rho}^{\mathrm{gas}}) S^{\mathrm{el}\dag} ] \label{eq:rhoprimeel}
\end{align}
and the $\nu\nu'$ matrix element is
\begin{align}
\hat{\rho}^{\prime \mathrm{el}}_{\nu\nu'} & = \mathrm{tr}_{\mathrm{gas}} S_{\nu\nu} [\hat{\rho}_{\nu\nu'}\otimes\hat{\rho}^{\mathrm{gas}}] S^\dag_{\nu'\nu'}\,\label{eq:rhoprimeelnunup}
\end{align}
since $S^{\mathrm{el}}_{\nu\nu'}=S_{\nu\nu'}$.  The inelastic term is
\begin{align}
\hat{\rho}^{\prime\mathrm{in}} & = \hat{\rho}^\prime - \hat{\rho}^{\prime\mathrm{el}}\,.
\end{align}
$\hat{\rho}^{\prime\mathrm{el}}$ corresponds to the portion of the ensemble which remains in the trap after the collision, while $\hat{\rho}^{\prime{\mathrm{in}}}$ describes particles ejected from the trap. 

The change of $\hat{\rho}$ in a single collision is 
\begin{align}
\Delta\hat{\rho} &= \hat{\rho}^\prime -\hat{\rho}
\end{align}
and we may perform a similar decomposition of $\Delta\hat{\rho}$ into terms due to elastic collisions,
\begin{align}
(\Delta\hat{\rho})^{\mathrm{el}} & = \hat{\rho}^{\prime\mathrm{el}}-\hat{\rho}
\end{align}
and inelastic collisions,
\begin{align}
(\Delta\hat{\rho})^{\mathrm{in}} & = \Delta\hat{\rho} - (\Delta\hat{\rho})^{\mathrm{el}} \,.
\end{align}

If we now consider the continuous-time evolution of the ensemble of trapped tracer molecules, the number of molecules in the trap will be continually decreasing due to trap loss.  Experimental measurements can only probe the molecules in the trap, hence we are interested in modelling only the portion of the original ensemble that remains in the trap.  The unnormalized density operator $\hat{\rho}^{\mathrm{el}}(t)$ describes the collection of molecules in the trap at time $t$.  Since all of the sample is trapped at $t=0$, $\hat{\rho}^{\mathrm{el}}(0)=\hat{\rho}(0)$.  The fraction of the initial sample still trapped at $t$ is $\mathrm{tr}\,\hat{\rho}^{\mathrm{el}}(t)$, where the trace is over the internal and translational degrees of freedom.  The molecules ejected from the trap cannot be realistically modelled, nor would it be of interest to do so since those molecules cannot be probed by experiment.  The ejected molecules vanish from the continually depleting trapped ensemble described by $\hat{\rho}^{\mathrm{el}}$ but beyond this their state is not defined.  The density operator $\hat{\rho}(t)$, which would ordinarily describe the state at $t$ of the entire ensemble of molecules, is therefore not well-defined and is not a quantity of physical interest since the outcome of all measurements on molecules in the trap is described by $\hat{\rho}^{\mathrm{el}}(t)$.

The reduced density matrix describing the internal state of the trapped ensemble is Hermitian and has matrix elements
$\rho^{\mathrm{el}}_{\nu\nu'}(t)=\Omega(2\pi\hbar)^{-3}\int d^3P\,\rho^{\mathrm{el}}_{\nu\nu'}(\gvec{P},\gvec{P};t)$.  These are complex numbers and should not be confused with the operators $\hat{\rho}^{\mathrm{el}}_{\nu\nu'}(t)$.  With this definition $\rho^{\mathrm{el}}_{\nu\nu'}(0)=\rho_{\nu\nu'}(0)$ (cf. Eq.~\eqref{eq:molinitialconds}).  

\subsection{Measures of coherence}
\label{sec:measures}

There are two quantities of interest for characterizing the degree of coherence between the internal states $\nu$ and $\nu'$ in the trapped ensemble.  The first is $|\rho^{\mathrm{el}}_{\nu\nu'}|$ and the second is 
\begin{align}
\eta_{\nu\nu'} & = \left(\frac{\rho^{\mathrm{el}}_{\nu\nu'}\rho^{\mathrm{el}}_{\nu'\nu}}{\rho^{\mathrm{el}}_{\nu\nu}\rho^{\mathrm{el}}_{\nu'\nu'}}\right)^{1/2} \label{eq:etadef}\,.
\end{align}
These are both nonnegative real numbers, and $0\leq \eta_{\nu\nu'}\leq1$.  For trapped molecules in a pure state $|\rho^{\mathrm{el}}_{\nu\nu'}|=\sqrt{\rho^{\mathrm{el}}_{\nu\nu}\rho^{\mathrm{el}}_{\nu'\nu'}}$ and 
 $\eta_{\nu\nu'}=1$. When there is a complete absence of coherence between the states $\nu$ and $\nu'$ both $|\rho^{\mathrm{el}}_{\nu\nu'}|$ and $\eta_{\nu\nu'}$ are 0.  Consider an experiment performed on the molecules in the trap which measures an observable $\hat{A}=\{A_{\nu\nu'}\}$ that involves only the internal but not the translational degrees of freedom.  The terms in the expectation value $\langle \hat{A} \rangle$ that depend on the coherence between the states $\nu$ and $\nu'$ are  $(\rho^{\mathrm{el}}_{\nu\nu'}A_{\nu'\nu}+\rho^{\mathrm{el}}_{\nu'\nu}A_{\nu\nu'})/\mathrm{tr}\,\rho^{\mathrm{el}}=2|\rho^{\mathrm{el}}_{\nu\nu'}||A_{\nu\nu'}|\cos\phi/\mathrm{tr}\,\rho^{\mathrm{el}}$ for some phase angle $\phi$.  Coherence in an experiment is often observed as temporal or spatial oscillations (i.e. interference fringes) in the value of some observable.  Hence, $\eta_{\nu\nu'}$ represents the ratio of the amplitude of observed interference fringes between $\nu$ and $\nu'$ to their maximum possible amplitude, which occurs when there is perfect coherence.  $|\rho_{\nu\nu'}^{\mathrm{el}}|$ is relevant for applications such as quantum computation where the size of the ensemble is an important aspect of the coherent signal.  Our goal is to determine $|\rho^{\mathrm{el}}_{\nu\nu'}(t)|$ and $\eta_{\nu\nu'}(t)$, and we proceed by first considering the time evolution of the density operator $\hat{\rho}^{\mathrm{el}}_{\nu\nu'}(t)$, which describes translational as well as internal degrees of freedom. 

\subsection{Master equation}

In a collision with incoming density matrix $\hat{\rho}^{\mathrm{el}}$, the change $\Delta \hat{\rho}^{\mathrm{el}}_{\nu\nu'}  = \hat{\rho}^{\prime\mathrm{el}}_{\nu\nu'}-\hat{\rho}^{\mathrm{el}}_{\nu\nu'}$ is
\begin{multline}
\Delta \hat{\rho}^{\mathrm{el}}_{\nu\nu'}  = \mathrm{tr}_{\mathrm{gas}}\left\{ \frac{i}{2}
[(T_{\nu\nu}+T^\dag_{\nu\nu})(\hat{\rho}^{\mathrm{el}}_{\nu\nu'}\otimes\hat{\rho}^{\mathrm{gas}})- (\hat{\rho}^{\mathrm{el}}_{\nu\nu'}\otimes\hat{\rho}^{\mathrm{gas}})   (T_{\nu'\nu'}+T^\dag_{\nu'\nu'})] \right. \label{eq:deltarho} \\
-\frac{1}{2}
\left.  \left[ \sum_{\nu''}
         T^\dag_{\nu\nu''}T_{\nu''\nu} (\hat{\rho}^{\mathrm{el}}_{\nu\nu'}\otimes\hat{\rho}^{\mathrm{gas}}) + (\hat{\rho}^{\mathrm{el}}_{\nu\nu'}\otimes\hat{\rho}^{\mathrm{gas}})T^\dag_{\nu'\nu''}T_{\nu''\nu'}
   \right] + T_{\nu\nu} (\hat{\rho}^{\mathrm{el}}_{\nu\nu'}\otimes\hat{\rho}^{\mathrm{gas}})
T^\dag_{\nu'\nu'}
\right\}\,,
\end{multline}
where we have used the relation
 \begin{equation}
i(T_{\nu\nu}-T^\dag_{\nu\nu}) = -\sum_{\nu''}T^\dag_{\nu''\nu}T_{\nu''\nu}\,,\label{eq:Tunitarity}
\end{equation}
which follows from the unitarity of the $S$ matrix.  

In the momentum representation Eq.~\eqref{eq:deltarho} takes the form
\begin{multline}
\Delta \rho^{\mathrm{el}}_{\nu\nu'}(\gvec{P},\gvec{P'}) = \frac{(2\pi\hbar)^3}{\Omega}
\left\{ 
 \rho^{\mathrm{el}}_{\nu\nu'}(\mathbf{P},\mathbf{P}') \int d^3\gvec{p}\,\mu(\gvec{p}) \left[ \frac{i}{2} 
\left( \braket{\mathrm{rel}(\gvec{p},\gvec{P})\nu|T_0+T^\dag_0|\mathrm{rel}(\gvec{p},\gvec{P})\nu} \right.\right.\right.
\\
 \shoveright
{\left.-\braket{\mathrm{rel}(\gvec{p},\gvec{P}')\nu'|T_0+T^\dag_0|\mathrm{rel}(\gvec{p},\gvec{P}')\nu'} \right)}
\\
 \mspace{72.0mu}
-\frac{1}{2}
\int d^3\gvec{Q}\,\sum_{\nu''}
\left(\left|\braket{\mathrm{rel}(\gvec{p}-\gvec{Q},\gvec{P}+\gvec{Q})\nu''|T_0|\mathrm{rel}(\gvec{p},\gvec{P})\nu}\right|^2\right.  \\
\shoveright{ \biggl.\left.
+\left|\braket{\mathrm{rel}(\gvec{p}-\gvec{Q},\gvec{P}'+\gvec{Q})\nu''|T_0|\mathrm{rel}(\gvec{p},\gvec{P}')\nu'}\right|^2 \right)\biggr]} \\
+ \int d^3\gvec{p}\,\mu(\gvec{p}) \int d^3\gvec{Q}\, \rho^{\mathrm{el}}_{\nu\nu'}\left(\gvec{P} -\gvec{Q},\gvec{P}'-\gvec{Q}\right) \braket{\mathrm{rel}(\gvec{p}-\gvec{Q},\gvec{P})\nu|T_0|\mathrm{rel}(\gvec{p},\gvec{P}-\gvec{Q})\nu} \\
 \left. \mspace{216.0mu}\times \braket{\mathrm{rel}(\gvec{p},\gvec{P}'-\gvec{Q})\nu'|T^\dag_0|\mathrm{rel}(\gvec{p}-\gvec{Q},\gvec{P}')\nu'}
\right\}\,.\label{eq:deltarhomomentum}
\end{multline}

Eq.~\eqref{eq:deltarhomomentum} contains terms of the form
\begin{equation}
\braket{\gvec{p}\nu|T_0+T^\dag_0|\gvec{p}\nu}=\frac{m_*}{2\pi\hbar}\delta(0)f_{\nu\nu}(\gvec{p},\gvec{p}) \label{eq:singularlinear}
\end{equation}
and
\begin{equation} 
|\braket{\gvec{p}'\nu''|T_0|\gvec{p}\nu}|^2=\frac{m_*^2}{(2\pi\hbar)^{2}}\delta^2(E'-E)|f_{\nu''\nu}(\gvec{p}',\gvec{p})|^2\label{eq:singularquadratic}
\end{equation}
which contain the undefined quantities $\delta(0)$ and $\delta^{2}(E'-E)$.  Hornberger and Sipe \cite{hs} present a normalization rule applicable to the single-collision expression Eq.~\eqref{eq:deltarhomomentum}.  However, we use an ultimately equivalent technique proposed by Adler \cite{adler} in which the normalization is carried out simultaneously with the passage from the single collision expression to a continuous-time equation for $\frac{\partial}{\partial t}\rho^{\mathrm{el}}(\gvec{P},\gvec{P}';t)$. 

According to Adler's method, on the left-hand side of Eq.~\eqref{eq:deltarhomomentum} we replace $\Delta \rho^{\mathrm{el}}_{\nu\nu'}(\gvec{P},\gvec{P'})$  with $\Delta \rho^{\mathrm{el}}_{\nu\nu'}(\gvec{P},\gvec{P'})=\rho^{\mathrm{el}}_{\nu\nu'}(\gvec{P},\gvec{P'};t+\Delta t) -\rho^{\mathrm{el}}_{\nu\nu'}(\gvec{P},\gvec{P'};t)$, which is the change in the density matrix due to collisions during a finite time interval $\Delta t$ with a bath ensemble of one particle.  The coarse-graining time $\Delta t$ must be longer than the duration of a collision but short compared to decoherence timescales \cite{hs,adler}.  The right-hand side is multiplied by $N$, the number of bath particles in the system volume $\Omega$.  

Adler shows that the square of the delta function can be written as $\delta^2(E'-E)=\delta(E'-E)\delta(0)$ and that the appropriate choice for $\delta(0)$ is $\delta(0)=\Delta t /(2\pi\hbar)$.  Hence $\delta^2(E'-E)=\delta(p'-p) \Delta t/(4\pi^2\hbar^2 p)$.  After dividing by $\Delta t$ one equates $\left[\rho^{\mathrm{el}}_{\nu\nu'}(\gvec{P},\gvec{P'};t+\Delta t)- \rho^{\mathrm{el}}_{\nu\nu'}(\gvec{P},\gvec{P'};t)\right]/\Delta t = \left[\frac{\partial}{\partial t}\rho^{\mathrm{el}}_{\nu\nu'}(\gvec{P},\gvec{P'};t)\right]_{\mathrm{coll}}$, the collisional contribution to
\begin{align}
\frac{\partial}{\partial t}\rho^{\mathrm{el}}_{\nu\nu}(\gvec{P},\gvec{P'};t) & = -\frac{i}{\hbar}\left( \frac{P^2}{2M}+\epsilon_\nu-\frac{P^{\prime 2}}{2M}-\epsilon_{\nu'} \right)\rho^{\mathrm{el}}_{\nu\nu}(\gvec{P},\gvec{P'};t)+\left[\frac{\partial}{\partial t}\rho^{\mathrm{el}}_{\nu\nu}(\gvec{P},\gvec{P'};t)\right]_{\mathrm{coll}}\,,
\end{align}
with the other term arising from free Hamiltonian evolution. 

Applying these manipulations to Eq.~\eqref{eq:deltarhomomentum} thus converts, on the left-hand side, $\Delta \rho^{\mathrm{el}}_{\nu\nu'}(\gvec{P},\gvec{P'})\longrightarrow \left[\frac{\partial}{\partial t}\rho^{\mathrm{el}}_{\nu\nu'}(\gvec{P},\gvec{P'};t)\right]_{\mathrm{coll}}$. 
On the the right-hand side the effected conversions are
\begin{align}
N\frac{(2\pi\hbar)^3}{\Omega} \braket{\mathbf{p}\nu|T_0+T^\dag_0|\mathbf{p}\nu} & \to 4\pi\hbar\frac{n_{\mathrm{gas}}}{m_*}\mathrm{Re}\,f_{\nu\nu}(\mathbf{p},\mathbf{p})  \label{eq:overallrule1} \\
N\frac{(2\pi\hbar)^3}{\Omega} \left|\braket{\mathbf{p}'\nu''|T_0|\mathbf{p}\nu}\right|^2 & \to \frac{n_{\mathrm{gas}}}{m_*p'} \delta\left(p'-\sqrt{p^2+2m_*(\epsilon_\nu-\epsilon_{\nu'')}}\right)|f_{\nu''\nu}(\gvec{p'},\gvec{p})|^2\,, \label{eq:overallrule2}
\end{align}
where $n_{\mathrm{gas}}= N/\Omega$. 

This leaves the final term on the right-hand side of Eq.~\eqref{eq:deltarhomomentum}, which contains the expression
$\braket{\mathrm{rel}(\gvec{p}-\gvec{Q},\gvec{P})\nu|T_0|\mathrm{rel}(\gvec{p},\gvec{P}-\gvec{Q})\nu} \braket{\mathrm{rel}(\gvec{p},\gvec{P}'-\gvec{Q})\nu'|T^\dag_0|\mathrm{rel}(\gvec{p}-\gvec{Q},\gvec{P}')\nu'}$ in which the initial momenta are different for the two matrix elements.  For the decoherence of internal state superpositions we are interested in the reduced density matrix for internal states having matrix elements $\rho^{\mathrm{el}}_{\nu\nu'}(t)= \int d^3P\,g_{\nu\nu'}(\gvec{P},t)$ where
$g_{\nu\nu'}(\gvec{P},t)=(2\pi\hbar)^{-3}\Omega \rho^{\mathrm{el}}_{\nu\nu'}(\gvec{P},\gvec{P};t)$.  Eq.~\eqref{eq:deltarhomomentum} established that evolution of the elements of the main diagonal
$g_{\nu\nu'}(\gvec{P};t)$ is independent of the off-diagonal elements $\rho^{\mathrm{el}}_{\nu\nu'}(\gvec{P},\gvec{P}';t)$, $\gvec{P}\neq\gvec{P}'$.  It is therefore sufficient to consider the restriction of Eq.~\eqref{eq:deltarhomomentum} to the $\gvec{P}=\gvec{P}'$ case, thereby avoiding concerns about the extension of Eq.~\eqref{eq:overallrule2} to off-diagonal terms.  With straightforward changes of variable, the resulting equation for full evolution is
\begin{align}
\frac{\partial}{\partial t}g_{\nu\nu'}(\gvec{P},t) & = \frac{i}{\hbar}(\epsilon_{\nu'}-\epsilon_\nu)g_{\nu\nu'}(\gvec{P},t) + \frac{n_{\mathrm{gas}}}{m_*}\left(1+r\right)^3 \left\{ g_{\nu\nu'}(\gvec{P},t) \int d^3p\,\mu\left( (1+r)\gvec{p}+r\gvec{P}\right) \right. \notag \\
& \qquad \qquad \qquad\times
 \left[ 2\pi\hbar i \mathrm{Re}\bigl(f_{\nu\nu}(\gvec{p},\gvec{p})-f_{\nu'\nu'}(\gvec{p},\gvec{p})\bigr)
 - \frac{p}{2}\left(\sigma^{\mathrm{tot}}_{\nu}(\gvec{p})+\sigma^{\mathrm{tot}}_{\nu'}(\gvec{p})\right) \right]  \notag \\
& \left.
+ \int d^3p\,d^2\uvec{n}\, \mu\left(r\left(\gvec{P}+p\uvec{n}\right)+\gvec{p}\right) g_{\nu\nu'}\left( \gvec{P}-\gvec{p}+p\uvec{n},t \right) p f_{\nu\nu}(p\uvec{n},\gvec{p})f^*_{\nu'\nu'}(p\uvec{n},\gvec{p})
 \right\}\,,\label{eq:gevolution}
\end{align}
where we have defined $r=m/M$.  The total cross section for scattering in state $\nu$ with pairwise relative momentum $\gvec{p}$,
\begin{align}
\sigma^{\mathrm{tot}}_\nu(\gvec{p}) & = \int d^2\uvec{n}\sum_{\nu''}
\frac{\sqrt{p^2+2m_*(\epsilon_{\nu}-\epsilon_{\nu''})}}{p}
\left|f_{\nu''\nu}\left(\sqrt{p^2+2m_*(\epsilon_{\nu}-\epsilon_{\nu''})}\uvec{n},\gvec{p}\right)\right|^2\,,
\end{align}
is related to the scattering amplitude in the forward direction by the optical theorem $\sigma^{\mathrm{tot}}_\nu(\gvec{p}) =(4\pi\hbar/p) \mathrm{Im}\,f_{\nu\nu}(\gvec{p},\gvec{p})$, which has been used in Eq.~\eqref{eq:gevolution}.

 The evolution of $\hat{\rho}^{\mathrm{el}}$ is not trace-preserving because of trap loss in inelastic collisions.  Hence there is no equation in Lindblad form for $\hat{\rho}^{\mathrm{el}}$ such that Eq.~\eqref{eq:gevolution}  can be obtained as its $(\gvec{P}\nu,\gvec{P}\nu')$ matrix element.

With $s$-wave scattering only, $f_{\nu\nu}(\gvec{p}',\gvec{p})= f_{\nu\nu}(p)$.  At low momenta the $s$-wave scattering amplitude may be expanded as
\begin{equation} 
f_{\nu\nu}(p) 
= -a_\nu+b_\nu p + c_\nu p^2+\ldots  = -(\alpha_\nu-i\beta_\nu)+(b^{\mathrm{r}}_\nu+ib^{\mathrm{i}}_\nu)p+\ldots \,, \label{eq:fexpansion}
\end{equation}
where the coefficients are in general complex and $a_\nu$ is the complex $s$-wave scattering length for particles in state $\nu$ \cite{complexa}. The sign of $a_\nu$ and the notation $\alpha_\nu$ and $-\beta_\nu$ for the real and imaginary parts of the scattering length are conventional.  

The elastic cross section is $\sigma^{\mathrm{el}}_{\nu}(\gvec{p})=\int d^2\uvec{n}\,|f_{\nu\nu}(p\uvec{n},\gvec{p})|^2$ and the inelastic cross section
$\sigma^{\mathrm{in}}_\nu(\gvec{p}) = \sigma^{\mathrm{tot}}_\nu(\gvec{p}) - \sigma^{\mathrm{el}}_\nu(\gvec{p})$. At leading orders in $p$, with the assumption of $s$-wave scattering only, 
\begin{align}
\sigma^{\mathrm{tot}}_\nu(p) & = \frac{4\pi\hbar}{p}(\beta_\nu+b^{\mathrm{i}}_\nu p+c^{\mathrm{i}}_\nu p^2+\ldots) \\
\sigma^{\mathrm{el}}_\nu(p) & = 4\pi\left[ |a_\nu|^2 + (-a_\nu b^*_\nu- a^*_\nu b_\nu)p+\ldots\right] \label{eq:sigmael} \\
\sigma^{\mathrm{in}}_\nu(p) & = 4\pi\left(\hbar\frac{\beta_\nu}{p} + \hbar b^{\mathrm{i}}_\nu-|a_\nu|^2 +\ldots \right) \label{eq:sigmain}  \,.
\end{align}
Because a cross section cannot be negative, $\beta_\nu \geq 0$.  

We will also use the expansion
\begin{align}
|f_{\nu\nu}(p)-f_{\nu'\nu'}(p)|^2 & = |a_{\nu}-a_{\nu'}|^2
-2\mathrm{Re}\left[(a_\nu-a_{\nu'})(b^*_\nu-b^*_{\nu'})\right]p \notag \\
& \qquad
+\left\{|b_\nu-b_{\nu'}|^2-2\mathrm{Re}[(a_\nu-a_{\nu'})(c^*_\nu-c^*_{\nu'})]\right\}p^2+\ldots  \,. \label{eq:ffexpand}
\end{align}

\section{Temperature dependence of decoherence}
\label{sec:tempdependence}

At this point it is convenient to  introduce dimensionless variables.  We introduce a characteristic length $l$, which we leave unspecified, that will drop out before the final results.  We define a dimensionless temperature $\theta=2mk_BTl^2/\hbar^2$,  $\gvec{Q}=\gvec{P}l/(\hbar\sqrt{\theta})$, $\tau=t\hbar n_{\mathrm{gas}}l/m$ and $\gamma_{\nu\nu'}(\gvec{Q},\tau) = (\hbar\sqrt{\theta}/l)^3g_{\nu\nu'}\left(\gvec{P}\hbar\sqrt{\theta}/l,\tau[\hbar n_{\mathrm{gas}}l/m]^{-1}\right)$.  These scalings have been chosen so that the independent variables describing the particle masses are $m$ and $r=m/M$, a property that will be used in Sec.~\ref{sec:smallr}.  

In the scaled variables Eq.~\eqref{eq:gevolution} is
\begin{align}
\frac{d}{d\tau}\gamma_{\nu\nu'}(\tau) & = G[\gamma_{\nu\nu'}(\tau)] \label{eq:gammaevolution}
\end{align}
where we have suppressed the $\gvec{Q}$-argument thereby indicating that we are considering the function $\gamma(\tau)=\gamma(\cdot,\tau)$ as an object which may be distinguished from its value evaluated at $\gvec{Q}$, $\gamma(\gvec{Q},\tau)$.  The linear operator $G$ acts on a function $h(\gvec{Q})$ as
\begin{multline}
G[h](\gvec{Q}) = i\frac{(\epsilon_\nu-\epsilon_{\nu'})m}{\hbar^2 n_{\mathrm{gas}}l} h(\gvec{Q})+
(1+r)^4 \\ \times\left\{ h(\gvec{Q}) \int d^3q\,2\pi i\left[\frac{f_{\nu\nu}\left(\frac{\hbar\theta^{1/2}}{l}\gvec{q},\frac{\hbar\theta^{1/2}}{l}\gvec{q}\right)-f^*_{\nu'\nu'}\left(\frac{\hbar\theta^{1/2}}{l}\gvec{q},\frac{\hbar\theta^{1/2}}{l}\gvec{q}\right)}{l}\right]\frac{e^{-[r\gvec{Q}+(1+r)\gvec{q}]^2}}{\pi^{3/2}}\right. \\
+
\theta^{1/2}\left.\int d^3q\, d^2\uvec{n}\, \frac{f_{\nu\nu}\left(\frac{\hbar\theta^{1/2}}{l}q\uvec{n},\frac{\hbar\theta^{1/2}}{l}\gvec{q}\right)f^*_{\nu'\nu'}\left(  \frac{\hbar\theta^{1/2}}{l}q\uvec{n},\frac{\hbar\theta^{1/2}}{l}\gvec{q}\right)}{l^2} h(\gvec{Q}-\gvec{q}+q\uvec{n})q\frac{e^{-(r\gvec{Q}+rq\uvec{n}+\gvec{q})^2}}{\pi^{3/2}}
\right\}\,.\label{eq:Gdef}
\end{multline}
$G$ depends on $\nu$ and $\nu'$ but we suppress this in the notation for clarity.  The initial conditions implied by Eq.~\eqref{eq:molinitialconds} are 
\begin{equation}
\gamma(\gvec{Q},0)=\rho_{\nu\nu'}(0)\left(\frac{r}{\pi}\right)^{3/2}e^{-rQ^2}\,.\label{eq:ic}
\end{equation}

Applying the assumption that there is only $s$-wave scattering, we substitute Eq.~\eqref{eq:fexpansion} into Eq.~\eqref{eq:Gdef}. 
Since $\gvec{p}=\hbar\sqrt{\theta}\gvec{q}/l$, we obtain an expansion in powers of $\theta^{1/2}$:
\begin{align}
G & = G_0 +\theta^{1/2}G_1 + \theta G_2 + \ldots\,, \label{eq:Gexpansion}
\end{align}
where
\begin{align}
G_0[h](\gvec{Q}) & = \kappa^{\nu\nu'}_0h(\gvec{Q})
\end{align}
with 
\begin{align}
\kappa^{\nu\nu'}_0 & = i\frac{(\epsilon_\nu-\epsilon_{\nu'})m}{\hbar^2 n_{\mathrm{gas}}l}-\frac{2\pi i(a_\nu-a^*_{\nu'})(1+r)}{l} 
\end{align}
and
\begin{multline}
G_1[h](\gvec{Q})  = (1+r)^{4} \left\{   2\pi \frac{i\hbar(b_\nu-b^*_{\nu'})}{l^2}  h(\gvec{Q}) \int d^3q\,q \frac{e^{-[r\gvec{Q}+(1+r)\gvec{q}]^2}}{\pi ^{3/2}} \right.  \\
 \qquad + \left. \frac{a_\nu a^*_{\nu'}}{l^2} \int d^3q\,d^2\uvec{n}\,qh(\gvec{Q}-\gvec{q}+q\uvec{n})\frac{e^{-(r\gvec{Q}+rq\uvec{n}+\gvec{q})^2}}{\pi^{3/2}}\right\}\,,\label{eq:G1def}
\end{multline}
and 
\begin{align}
G_2[h](\gvec{Q}) & =
(1+r)^{4} \left\{ h(\gvec{Q}) 2\pi i \frac{\hbar^2(c_\nu-c^*_{\nu'})}{l^3}\int d^3q\,q^2\frac{e^{-[r\gvec{Q}+(1+r)\gvec{q}]^2}}{\pi^{3/2}} \right.  \notag \\
& \qquad \qquad
\left.+\frac{\hbar(-a_\nu b^*_{\nu'}-b_\nu a^*_{\nu'})}{l^3} \int d^3q\,d^2\uvec{n}\,h(\gvec{Q}-\gvec{q}+q\uvec{n})q^2 \frac{e^{-(r\gvec{Q}+rq\uvec{n}+\gvec{q})^2}}{\pi^{3/2}}
\right\}\,.
\label{eq:G2def}
\end{align}

For notational brevity we define $\int h \equiv \int d^3Q\,h(\gvec{Q})$  for any function $h(\gvec{Q})$.   The zeroth-order truncation of Eq.~\eqref{eq:gammaevolution} has solution $\gamma_{\nu\nu'}(\tau)=\gamma_{\nu\nu'}(0)e^{\kappa^{\nu\nu'}_0\tau}$ and yields $\rho_{\nu\nu'}(\tau)=\rho_{\nu\nu'}(0)e^{\kappa^{\nu\nu'}_0\tau}$.  
Observe that 
\begin{equation}
\mathrm{Re}\,\kappa^{\nu\nu'}_0  = -2\pi(\beta_\nu+\beta_{\nu'})/l \leq 0\,,\label{eq:rez1lezero}
\end{equation}
hence $e^{\kappa^{\nu\nu'}_0\tau}$ is a decaying term except when equality holds, which occurs when the leading order of inelastic scattering vanishes for both internal states.  
The solution of Eq.~\eqref{eq:gammaevolution} has the form $\gamma_{\nu\nu'}(\tau)=e^{\kappa^{\nu\nu'}_0\tau}F_{\nu\nu'}(\tau)$, where
\begin{equation}
\frac{d}{d\tau}F_{\nu\nu'}(\tau)=\theta^{1/2}G_1[F_{\nu\nu'}]+\theta G_2[F_{\nu\nu'}]+\ldots\,.\label{eq:Feqn}
\end{equation}
$F_{\nu\nu'}(\tau)$ in general does not describe simple exponential decay.  Even if we truncate the right-hand side at order $\theta^{1/2}$, Eq.~\eqref{eq:Feqn} is difficult to solve because of the complicated form of $G_1$. Instead, we adopt a perturbation approach to the analysis of Eq.~\eqref{eq:gammaevolution}, expanding
\begin{equation} 
\gamma_{\nu\nu'} = \gamma^{\nu\nu'}_0 + \theta^{1/2} \gamma^{\nu\nu'}_1 + \theta \gamma^{\nu\nu'}_2 + \ldots \label{eq:gammaexpansion}\,.
\end{equation}
Substituting this and Eq.~\eqref{eq:Gexpansion} into Eq.~\eqref{eq:gammaevolution}, we obtain
\begin{equation}
\frac{d}{d\tau}\gamma^{\nu\nu'}_0+\theta^{1/2}\frac{d}{d\tau}\gamma^{\nu\nu'}_1 +\ldots = G_0[\gamma^{\nu\nu'}_0]+\theta^{1/2}(G_0[\gamma^{\nu\nu'}_1]+G_1[\gamma^{\nu\nu'}_0]) +\ldots\,.\label{eq:perturbation}
\end{equation}
The initial conditions are $\gamma^{\nu\nu'}_0(0)=\gamma_{\nu\nu'}(0)$ and $\gamma^{\nu\nu'}_k(0)=0$ for $k\geq 1$.  Equating coefficients of like powers of $\theta^{1/2}$ and solving at the lowest three orders we obtain
\begin{align}
\gamma^{\nu\nu'}_0(\tau) & = e^{\kappa^{\nu\nu'}_0\tau}\gamma_{\nu\nu'}(0) \label{eq:gamma0soln} \\
\gamma^{\nu\nu'}_1(\tau) & = \tau e^{\kappa^{\nu\nu'}_0\tau}G_1[\gamma_{\nu\nu'}(0)]\label{eq:gamma1soln} \\ 
\gamma^{\nu\nu'}_2(\tau) & = e^{\kappa^{\nu\nu'}_0\tau}\left(\tau G_2[\gamma_{\nu\nu'}(0)] +\frac{1}{2}\tau^2G_1[G_1[\gamma_{\nu\nu'}(0)]]\right)\,. \label{eq:gamma2soln}
\end{align}
These are the leading terms in the expansion of $e^{G\tau}[\gamma_{\nu\nu'}(0)]$, which is the formal solution to Eq.~\eqref{eq:gammaevolution}.

We substitute Eqs.~\eqref{eq:gamma0soln} and \eqref{eq:gamma1soln} into Eq.~\eqref{eq:gammaexpansion}, and integrate over $\gvec{Q}$.  The integrals in the resulting expression are
\begin{align}
\int G_1[\gamma^{\nu\nu'}_0(0)] & = \frac{2(1+r)^{1/2}}{\pi^{1/2}}\left(\frac{2\pi i\hbar(b_\nu-b^*_{\nu'})+4\pi a_\nu a^*_{\nu'}}{l^2}\right)\rho_{\nu\nu'}(0) 
 \label{eq:inteval1} \\
\int G_2[\gamma^{\nu\nu'}_0(0)] & =  \frac{ 3\left[2\pi i\hbar^2(c_\nu-c^*_{\nu'})- 4\pi \hbar(a_\nu b^*_{\nu'} +b_\nu a^*_{\nu'})\right]}{2l^3} \rho_{\nu\nu'}(0)    \label{eq:inteval21} \\
\int G_1[G_1[\gamma^{\nu\nu'}_0(0)]] & =  \frac{1}{\pi}\left[\frac{2\pi i \hbar (b_\nu-b^*_{\nu'})+4\pi a_\nu a^*_{\nu'}}{l^2}\right]^2 \notag \\ & \qquad \times
\left[3(2r+1)^{1/2}+\frac{1+2r+3r^2}{r}\sin^{-1}\frac{r}{r+1}\right]
\rho_{\nu\nu'}(0) \label{eq:inteval22} \,.
\end{align}
The evaluation of Eq.~\eqref{eq:inteval22} is discussed in the Appendix.

Converting back to unscaled variables and using Eqs.~\eqref{eq:inteval1}-\eqref{eq:inteval22}, we obtain
\begin{equation}
\rho^{\mathrm{el}}_{\nu\nu'}(t)  = e^{z^{\nu\nu'}_0t}\rho_{\nu\nu'}(0)\left[1+T^{1/2} z^{\nu\nu'}_1 t  + T \left(z^{\nu\nu'}_{2,1}t+ z^{\nu\nu'}_{2,2}\frac{t^2}{2} \right) + \ldots \right] \,, \label{eq:rhoversust} 
\end{equation}
where
\begin{align}
z^{\nu\nu'}_0 & = \frac{i(\epsilon_{\nu'}-\epsilon_\nu)}{\hbar}-\frac{2\pi i (a_\nu-a^*_{\nu'})\hbar n_{\mathrm{gas}}}{m_*} \\
z^{\nu\nu'}_1 & =  \frac{2^{5/2}\pi^{1/2}k_B^{1/2}n_{\mathrm{gas}}}{m_*^{1/2}}[i\hbar(b_\nu-b^*_{\nu'})+2 a_\nu a^*_{\nu'}]\\
z^{\nu\nu'}_{2,1} & = 6\pi n_{\mathrm{gas}}k_B r^{3/2}[i\hbar(c_\nu-c^*_{\nu'})- 2(a_\nu b^*_{\nu'} +b_\nu a^*_{\nu'})]
\end{align}
and
\begin{align}
z^{\nu\nu'}_{2,2} & = 
\frac{8 \pi k_Bn_{\mathrm{gas}}^2}{m}\left[3(2r+1)^{1/2}+\frac{1+2r+3r^2}{r}\sin^{-1}\frac{r}{r+1}\right]\left[ i \hbar (b_\nu-b^*_{\nu'}) + 2 a_\nu a^*_{\nu'} \right]^2
\,.
\end{align}
Note that
\begin{equation}
\mathrm{Re}\,z^{\nu\nu'}_0 = -\frac{2\pi\hbar n_{\mathrm{gas}}}{m_*}(\beta_{\nu}+\beta_{\nu'})\leq 0 \label{eq:Rez0}\,.
\end{equation}
Referring to Eq.~\eqref{eq:sigmain}, we observe that
equality occurs when the coefficient of the $p^{-1}$ term vanishes in both $\sigma^{\mathrm{in}}_{\nu}(p)$ and $\sigma^{\mathrm{in}}_{\nu'}(p)$.
As well,
\begin{equation}
\mathrm{Re}\,z^{\nu\nu'}_1 = \frac{2^{5/2}\pi^{1/2}k_B^{1/2}n_{\mathrm{gas}}}{m_*^{1/2}}\left(-\hbar b^{\mathrm{i}}_{\nu}+|a_\nu|^2 - \hbar b^{\mathrm{i}}_{\nu'} + |a_{\nu'}|^2-|a_{\nu}-a_{\nu'}|^2\right)\,. \label{eq:Rez1}
\end{equation}
Note from Eq.~\eqref{eq:sigmain} that $\hbar b^{\mathrm{i}}_{\nu}-|a_\nu|^2$ is the coefficient of the $p$-independent term of $\sigma_{\nu}^{\mathrm{in}}(p)$.  If these coefficients are nonegative for $\sigma_\nu^{\mathrm{in}}$ and $\sigma_{\nu'}^{\mathrm{in}}$ then $\mathrm{Re}\,z^{\nu\nu'}_1\leq0$.  The nonnegativity of $\sigma^{\mathrm{in}}$ is not sufficient to establish the nonnegativity of the $p$-independent coefficient.  On physical grounds the inequality is expected to hold. 

\subsection{Total coherent signal}

The first of the two measures of coherence between the states $\ket{\nu}$ and $\ket{\nu'}$ we calculate is
\begin{align}
|\rho^{\mathrm{el}}_{\nu\nu'}(t)| & = \sqrt{\rho^{\mathrm{el}}_{\nu\nu'}(t) \rho^{\mathrm{el}}_{\nu'\nu}(t)}\,.
\end{align}
From Eq.~\eqref{eq:rhoversust} and the small-$|x|$ expansion $(1+x)^{1/2}=1+x/2-x^2/8+\ldots$ we find 
\begin{align}
|\rho^{\mathrm{el}}_{\nu\nu'}(t)| & = |\rho_{\nu\nu'}(0)|e^{-\zeta^{\nu\nu'}_0t}\left[1+\zeta^{\nu\nu'}_{1}T^{1/2}t+T\left( \zeta^{\nu\nu'}_{2,1}t +\zeta^{\nu\nu'}_{2,2}\frac{t^2}{2}
\right) + \ldots\right] \label{eq:absrhoversust}
\end{align}
with
\begin{align}
\zeta_0 & = -\mathrm{Re}\,z^{\nu\nu'}_0 \\
\zeta_1 & = \mathrm{Re}\,z^{\nu\nu'}_1 \\
\zeta_{2,1} & = \mathrm{Re}\,z^{\nu\nu'}_{2,1} 
\end{align}
and 
\begin{align}
\zeta_{2,2} & = \mathrm{Re}\,z^{\nu\nu'}_{2,2}+\left(\mathrm{Im}\,z^{\nu\nu'}_1\right)^2 
\,.
\end{align}
Differentiating Eq.~\eqref{eq:absrhoversust} gives
\begin{align}
\frac{d}{dt}|\rho^{\mathrm{el}}_{\nu\nu'}(t)| & = |\rho_{\nu\nu'}(0)|e^{-\zeta_{0}t} \notag \\
& \qquad \times \left\{
-\zeta_{0}+ \zeta_{1}(1-\zeta_{0}t)T^{1/2}+T\left[
\zeta_{2,1}(1-\zeta_{0}t)+\zeta_{2,2}\left(t-\frac{\zeta_{0}t^2}{2}\right)
\right]+\ldots
\right\}\,.\label{eq:ddtabsrho}
\end{align}
The $T$-independent leading term is the result of trap loss from inelastic collisions.  

\begin{figure}[hbtp]
\includegraphics{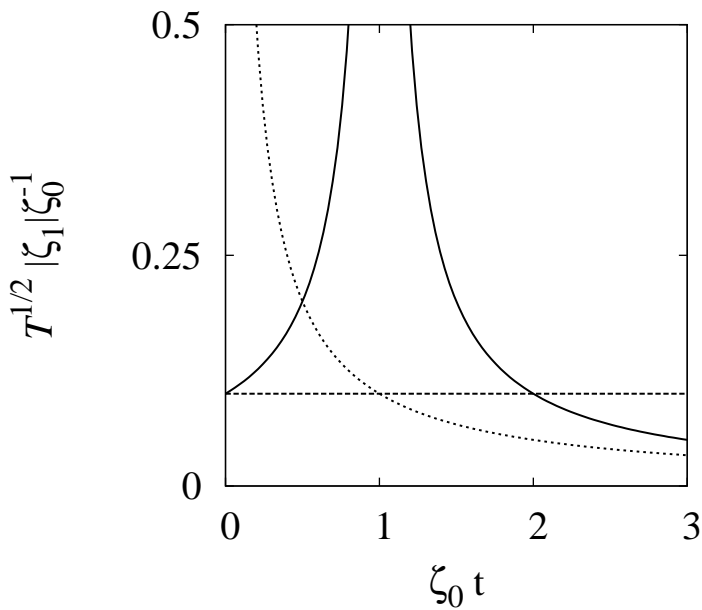}
\caption{\label{fig1}The curves are $|\zeta_{1}||1-\zeta_{0}t|T^{1/2} = 0.1\zeta_{0}$ (solid), $T^{1/2} =0.1 \zeta_0 / |\zeta_1|$ (dashed) and  $t =0.1(|\zeta_1|T^{1/2})^{-1}$ (dotted).}
\end{figure}

It is interesting to determine when the $T^{1/2}$-dependent term may be neglected in comparison with the $T$-independent term.  This occurs when 
\begin{align}
|\zeta_{1}(1-\zeta_{0}t)|T^{1/2} & \ll \zeta_{0}\,. \label{eq:zerothordercondition}
\end{align}
This condition is met for a time interval beginning at $t=0$ when 
\begin{align}
T^{1/2} & \ll \frac{\zeta_0}{|\zeta_1|} \label{eq:Tcondition}
\end{align}
and remains valid while 
\begin{align}
t & \ll \frac{1}{|\zeta_1|T^{1/2}}\,. \label{eq:cutofftime}
\end{align}
The regions of the $(t,T^{1/2})$ space satisfying Eqs.~\eqref{eq:zerothordercondition}-\eqref{eq:cutofftime} are shown in Fig.~\ref{fig1}.

It is noteworthy that the $T^{1/2}$ term becomes significant given sufficiently long time rather than remaining negligible indefinitely.   Eq.~\eqref{eq:gammaevolution} does not describe simple exponential decay.  In the perturbation formulation we may attribute the increasing $\theta^{1/2}$ term to the cumulative effect of the $\theta^{1/2}$ component $G_1$ of the time evolution operator $G$. The $\theta$-independent component $G_0$ of the operator $G$ of Eq.~\eqref{eq:gammaevolution} causes the pseudodistribution $\gamma_{\nu\nu'}(\gvec{Q})$ to be scaled by a $\gvec{Q}$-independent factor, but there is no change in its shape.  $G_1$ contains two terms (see Eq.~\eqref{eq:G1def}), the first of which corresponds to $\gvec{Q}$-dependent decay.  The second term describes $\gvec{Q}$-dependent decay and also the ``redistribution'' of $\gamma_{\nu\nu'}(\gvec{Q})$ between different $\gvec{Q}$ values due to thermalization, indicated by the $\gvec{Q}$-dependent convolution appearing in the term.  These effects change the shape of $\gamma_{\nu\nu'}(\gvec{Q})$ as opposed to scaling it by a constant. Their cumulative contribution will be negligible at short times but significant at sufficiently long times.

\subsection{Relative coherence in trapped sample}

The time-dependent relative coherence between the states $\ket{\nu}$ and $\ket{\nu'}$ in the trapped sample is found by substituting Eq.~\eqref{eq:rhoversust} into Eq.~\eqref{eq:etadef}.  We use the small-$|x|$ expansions of $(1+x)^{1/2}$ and $(1+x)^{-1/2}=1-x/2+3x^2/8+\ldots$, and observe that $(z_0^{\nu\nu'}+z_0^{\nu'\nu}-z_0^{\nu\nu}-z_0^{\nu'\nu'})/2=0$ and $\rho_{\nu\nu'}^{1/2}(0) \rho_{\nu'\nu}^{1/2}(0) \rho_{\nu\nu}^{-1/2}(0) \rho_{\nu'\nu'}^{-1/2}(0)=\eta_{\nu\nu'}(0)$.  We obtain
\begin{align}
\eta_{\nu\nu'}(t) = \eta_{\nu\nu'}(0)\left[1+T^{1/2}t\xi_1 + T\left(\xi_{2,1}t +\frac{t^2}{2}\xi_{2,2} \right) + \ldots\right] \label{eq:etaversust}
\end{align}
where the real-valued coefficients
\begin{align}
\xi_1 & = \frac{z_1^{\nu\nu'}+z_1^{\nu\nu'}-z_1^{\nu\nu}-z_1^{\nu'\nu'}}{2} 
\notag \\
& = -\frac{2^{5/2}\pi^{3/2}n_{\mathrm{gas}}k_B^{1/2}}{m_*^{1/2}}|a_\nu-a_{\nu'}|^2 \label{eq:xi1defn}\\
\xi_{2,1} & = \frac{z_{2,1}^{\nu\nu'}+z_{2,1}^{\nu\nu'}+z_{2,1}^{\nu\nu}-z_{2,1}^{\nu'\nu'}}{2} \notag \\
& = 12\pi n_{\mathrm{gas}}k_B r^{3/2}\mathrm{Re}[(a_\nu-a_{\nu'})(b_{\nu}^*-b_{\nu'}^*)] \label{eq:xi21defn}\\
\xi_{2,2} 
& = \frac{z_{2,2}^{\nu\nu'}+z_{2,2}^{\nu\nu'}-z_{2,2}^{\nu\nu}-z_{2,2}^{\nu'\nu'}}{2} \notag \\
& \qquad
+\frac{2\left[(z_1^{\nu\nu})^2 + (z_1^{\nu'\nu'})^2 -(z_1^{\nu\nu'})^2-(z_1^{\nu'\nu})^2\right]
+\left[z_1^{\nu\nu'}+z_1^{\nu'\nu}-\left(z_1^{\nu\nu}+z_1^{\nu'\nu'}\right)
\right]^2
}{4} \notag \\ 
& =
-\frac{8 \pi k_Bn_{\mathrm{gas}}^2}{m}\left[3(2r+1)^{1/2}+\frac{1+2r+3r^2}{r}\sin^{-1}\left(\frac{r}{r+1}\right)-4(1+r)\right] \notag \\
& \qquad\times\left\{ |\hbar(b_\nu-b_{\nu'})|^2 -4\mathrm{Im}\left[
\hbar(b_\nu a_\nu- b_{\nu'}a_{\nu'})(a_\nu^*-a_{\nu'}^*)\right] 
+2|a_\nu-a_{\nu'}|^2|a_\nu+a_{\nu'}|^2
\right\} \notag \\
&+ \frac{32\pi^3n_{\mathrm{gas}}^2k_B}{m_*}|a_\nu-a_{\nu'}|^4
\,. \label{eq:xi22defn}
\end{align}
$\xi_1\leq0$, while the signs of $\xi_{2,1}$ and $\xi_{2,2}$ are unconstrained.  Observe the appearance of coefficients from Eq.~\eqref{eq:ffexpand} in Eqs.~\eqref{eq:xi1defn} and \eqref{eq:xi21defn}.

\begin{figure}[hbtp]
\includegraphics{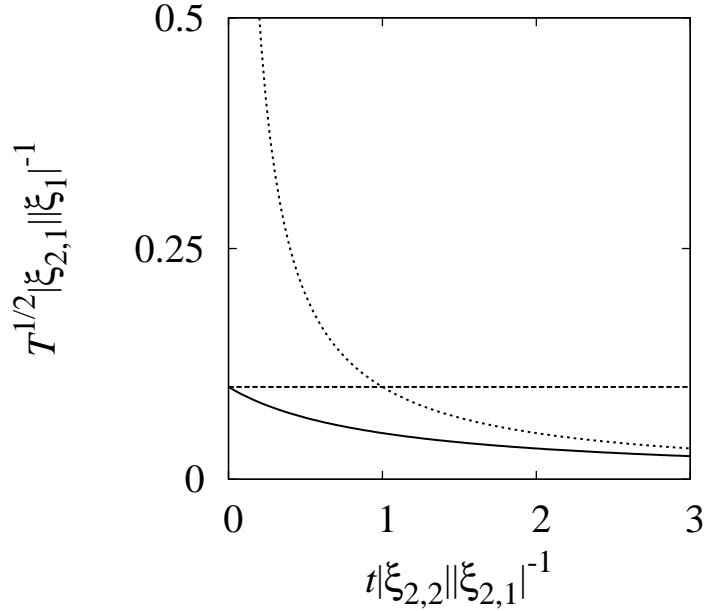}
\caption{\label{fig2}The curves are $T^{1/2}|\xi_{2,1} + t\xi_{2,2}| = 0.1 |\xi_1|$ (solid), $T^{1/2} =0.1|\xi_1|/|\xi_{2,1}|$ (dashed), and $t =0.1|\xi_1|/(|\xi_{2,2}|T^{1/2})$ (dotted).  The case depicted is that in which $\xi_{2,1}$ and $\xi_{2,2}$ have the same sign.}
\end{figure}

The decoherence rate is  
\begin{equation}
\frac{d}{dt}\eta_{\nu\nu'}(t)  = \eta_{\nu\nu'}(0) \left[T^{1/2}\xi_1+T(\xi_{2,1}+t\xi_{2,2})+\ldots\right]\,. \label{eq:ddteta}
\end{equation}
We wish to determine when the $T$-dependent contribution to decoherence as measured by $\eta_{\nu\nu'}(t)$ may be neglected by comparison with the $T^{1/2}$-dependent contribution.  The condition that must be satisfied is
\begin{equation}
 T^{1/2}|\xi_{2,1} + t\xi_{2,2}| \ll |\xi_1|\,,\label{eq:ineqcondition0}
\end{equation}
from which it follows that it is sufficient that both
\begin{equation}
T^{1/2} \ll \frac{|\xi_1|}{|\xi_{2,1}|} = \frac{2^{3/2}\pi^{3/2}|a_\nu-a_{\nu'}|^2}{3k_B^{1/2}m_*^{1/2}r^{3/2}|\mathrm{Re}\left[(a_\nu-a_{\nu'})(b_{\nu}^*-b_{\nu'}^*)\right]|} \label{eq:ineqcondition1}
\end{equation}
and
\begin{equation}
t \ll \frac{|\xi_1|}{|\xi_{2,2}|T^{1/2}}\,. \label{eq:ineqcondition2}
\end{equation}
With Eq.~\eqref{eq:ineqcondition1} holding, Eq.~\eqref{eq:ineqcondition2} is equivalent to the condition that $t$ be comparable to or smaller than $|\xi_{2,1}|/|\xi_{2,2}|$.  Fig.~\ref{fig2} illustrates the regions corresponding to Eqs.~\eqref{eq:ineqcondition0}-\eqref{eq:ineqcondition2} for the case where $\xi_{2,1}$ and $\xi_{2,1}$ have the same sign.  In obtaining Eq.~\eqref{eq:ineqcondition2}, unity has been neglected in comparison with $T^{1/2}|\xi_{2,2}|/|\xi_{2,1}|$ which results in a less restrictive condition, indicated in Fig.~\ref{fig2} by the solid curve lying above the dotted curve.  This is permissible as these conditions are order of magnitude estimates rather than strict inequalities.

\section{Light buffer gas particle}
\label{sec:smallr}

Helium is a likely choice for the buffer gas in experiments, so the low-$r$ regime, corresponding to a light buffer gas particle, is of particular interest.  Since there are two independent mass variables, the $r\to0^+$ limit may be approached along different paths in the two-dimensional space.  The behavior as $r\to0^+$ with $M$ held constant is highly singular.  We consider the limit as $r\to0^+$ with $m$ constant, which will allow us to obtain low-$r$ approximations and conditions for their validity.  The reduced variables introduced in Sec.~\ref{sec:tempdependence} were chosen to make $r$ and $m$ independent variables.

Beginning from Eq.~\eqref{eq:gammaevolution}, we make the low-$r$ expansion 
\begin{equation}
G=i\frac{(\epsilon_\nu-\epsilon_{\nu'})m}{\hbar^2 n_{\mathrm{gas}}l}+\tilde{G}_0+r\tilde{G}_1 +r^2\tilde{G}_2+\ldots\,,\label{eq:rexpansion}
\end{equation}
where for a function $h(\gvec{Q})$
\begin{multline}
\tilde{G}_k[h](\gvec{Q}) = h(\gvec{Q}) \int d^3q\, \frac{2\pi i \left[f_{\nu\nu}\left(\frac{\hbar\theta^{1/2}}{l}\gvec{q},\frac{\hbar\theta^{1/2}}{l}\gvec{q}\right)-f^*_{\nu'\nu'}\left(\frac{\hbar\theta^{1/2}}{l}\gvec{q},\frac{\hbar\theta^{1/2}}{l}\gvec{q}\right)\right]}{l}\frac{e^{-q^2}}{\pi^{3/2}}\mathcal{G}_k(\gvec{Q},\gvec{q},\uvec{q}) \\
+ \theta^{1/2}\int d^3q\,d^2\uvec{n}\,\frac{f_{\nu\nu}\left(\frac{\hbar\theta^{1/2}}{l}q\uvec{n},\frac{\hbar\theta^{1/2}}{l}\gvec{q}\right)f^*_{\nu'\nu'}\left(\frac{\hbar\theta^{1/2}}{l}q\uvec{n},\frac{\hbar\theta^{1/2}}{l}\gvec{q}\right)}{l^2}q\frac{e^{-q^2}}{\pi^{3/2}}\mathcal{G}_k(\gvec{Q},\gvec{q}, \uvec{n})h(\gvec{Q}-\gvec{q}+q\uvec{n})
\end{multline}
and
\begin{align}
\mathcal{G}_0(\gvec{Q},\gvec{q},\uvec{n}) & = 1 \\
\mathcal{G}_1(\gvec{Q},\gvec{q},\uvec{n}) & = 4-2\gvec{q}\cdot(\gvec{Q}+q\uvec{n})  \\
\mathcal{G}_2(\gvec{Q},\gvec{q},\uvec{n}) & = 6+[2\gvec{q}\cdot(\gvec{Q}+q\uvec{n})]^2-(\gvec{Q}+\gvec{q}\uvec{n})^2-8\gvec{q}\cdot(\gvec{Q}+q\uvec{n}) \,.
\end{align}
Due to the complicated structure of the operators $\tilde{G}_k$ a perturbation expansion in small $r$ does not lead to anything useful.

Inserting Eq.~\eqref{eq:rexpansion} into Eq.~\eqref{eq:gammaevolution} and integrating over $\gvec{Q}$ we obtain
\begin{align}
\frac{d}{d\tau}\rho^{\mathrm{el}}_{\nu\nu'}(\tau)  & = i\frac{(\epsilon_{\nu'}-\epsilon_{\nu})m}{\hbar^2 n_{\mathrm{gas}}l}\rho^{\mathrm{el}}_{\nu\nu'}(\tau) \notag \\ 
& \; +\int d^3Q\,\gamma_{\nu\nu'}(\gvec{Q},\tau) \int d^3q\,\frac{e^{-q^2}}{\pi^{3/2}}\left[\mathcal{G}_0(\gvec{Q},\gvec{q},\uvec{q})+r\mathcal{G}_1(\gvec{Q},\gvec{q},\uvec{q}) +r^2\mathcal{G}_2(\gvec{Q},\gvec{q},\uvec{q}) +\ldots \right] \notag \\
& \qquad \qquad \times\left\{
\frac{2\pi i \left[f_{\nu\nu}\left(\frac{\hbar\theta^{1/2}}{l}\gvec{q},\frac{\hbar\theta^{1/2}}{l}\gvec{q}\right)-f^*_{\nu'\nu'}\left(\frac{\hbar\theta^{1/2}}{l}\gvec{q},\frac{\hbar\theta^{1/2}}{l}\gvec{q}\right)\right]}{l} \right. \notag \\
& \qquad \qquad \qquad \left.
+q\theta^{1/2}\int d^2\uvec{n}\,
\frac{f_{\nu\nu}\left(\frac{\hbar\theta^{1/2}}{l}q\uvec{n},\frac{\hbar\theta^{1/2}}{l}\gvec{q}\right)f^*_{\nu'\nu'}\left(\frac{\hbar\theta^{1/2}}{l}q\uvec{n},\frac{\hbar\theta^{1/2}}{l}\gvec{q}\right)}{l^2}
\right\}\,.
\end{align}

If the quantity in braces is independent of $\uvec{q}$ then $\uvec{q}\cdot\uvec{Q}$ averages to zero over one of the angular integrations and the integrals of the terms containing $\gvec{Q}$ in $\mathcal{G}_k(\gvec{Q},\gvec{q},\uvec{q})$ vanish for $k=0$ and 1.  Hence the $\gvec{q}$ integral will be independent of $\gvec{Q}$.  This spherical symmetry condition is satisfied at ultracold temperatures.  In this case 
\begin{equation}
\frac{d}{d\tau}\rho^{\mathrm{el}}_{\nu\nu'}(\tau) = \left(i\frac{(\epsilon_{\nu'}-\epsilon_{\nu})m}{\hbar^2 n_{\mathrm{gas}}l} + w^{\nu\nu'}_0 \right)\rho^{\mathrm{el}}_{\nu\nu'}(\tau) + r w^{\nu\nu'}_1\rho^{\mathrm{el}}_{\nu\nu'}(\tau) +r^2\int\,\tilde{G}_2[\gamma_{\nu\nu'}(\tau)]+\ldots \label{eq:drhodtaurexpansion}
\end{equation}
where 
\begin{multline}
w_k = \int d^3q\,\frac{e^{-q^2}}{\pi^{3/2}}\left(4-2q^2\right)^{k}
\left\{
\frac{2\pi i \left[f_{\nu\nu}\left(\frac{\hbar\theta^{1/2}}{l}\gvec{q},\frac{\hbar\theta^{1/2}}{l}\gvec{q}\right)-f^*_{\nu'\nu'}\left(\frac{\hbar\theta^{1/2}}{l}\gvec{q},\frac{\hbar\theta^{1/2}}{l}\gvec{q}\right)\right]}{l} \right. \\
\left.
+q\theta^{1/2}\int d^2\uvec{n}\,
\frac{f_{\nu\nu}\left(\frac{\hbar\theta^{1/2}}{l}q\uvec{n},\frac{\hbar\theta^{1/2}}{l}\gvec{q}\right)f^*_{\nu'\nu'}\left(\frac{\hbar\theta^{1/2}}{l}q\uvec{n},\frac{\hbar\theta^{1/2}}{l}\gvec{q}\right)}{l^2}
\right\}\,.
\label{eq:wkdefs}
\end{multline}
The evolution of $\rho^{\mathrm{el}}_{\nu\nu'}(\tau)$ is exponential up to the first order in $r$.  

At ultracold temperatures we may use Eq.~\eqref{eq:fexpansion}.  We obtain the expansions
\begin{align}
w_0^{\nu\nu'} & = -2\pi\frac{i(a_{\nu}-a_{\nu'}^*)}{l} + \theta^{1/2} 
4\pi^{1/2}\frac{[i\hbar(b_{\nu}-b_{\nu'}^*)+2a_{\nu}a_{\nu'}^*]}{l^2} \notag \\
& \qquad\qquad\qquad\qquad\qquad\qquad
+\theta 3\pi\frac{[i\hbar^{2}(c_{\nu}-c_{\nu'}^*)-2\hbar(a_{\nu}b_{\nu'}^*+b_{\nu}a_{\nu'}^*)]}{l^3}+\ldots \label{eq:w0}\\
w_1^{\nu\nu'} & = -2\pi\frac{i(a_{\nu}-a_{\nu'}^*)}{l}-\theta 3\pi\frac{[i\hbar^{2}(c_{\nu}-c_{\nu'}^*)-2\hbar(a_{\nu}b_{\nu'}^*+b_{\nu}a_{\nu'}^*)]}{l^3}+\ldots \label{eq:w1}\,.
\end{align}

\subsection{Total coherent signal}
\label{sec:smallrabsrho}

Using the relationship
\begin{align}
\frac{d}{d\tau}|\rho^{\mathrm{el}}_{\nu\nu'}(\tau)| & = \frac{1}{2|\rho^{\mathrm{el}}_{\nu\nu'}(\tau)|}\left(\rho^{\mathrm{el}}_{\nu'\nu}(\tau)\frac{d}{dt}\rho^{\mathrm{el}}_{\nu\nu'}(\tau) +  \rho^{\mathrm{el}}_{\nu\nu'}(\tau)\frac{d}{dt}\rho^{\mathrm{el}}_{\nu'\nu}(\tau) \right)
\end{align}
and Eq.~\eqref{eq:drhodtaurexpansion} we obtain
\begin{align}
\frac{d}{d\tau}|\rho^{\mathrm{el}}_{\nu\nu'}(\tau)| & = \mathrm{Re}\,w_0^{\nu\nu'}|\rho^{\mathrm{el}}_{\nu\nu'}|
+r\mathrm{Re}\,w_1^{\nu\nu'}|\rho^{\mathrm{el}}_{\nu\nu'}|
+\frac{r^2}{|\rho^{\mathrm{el}}_{\nu\nu'}|}\mathrm{Re}\left(  \rho^{\mathrm{el}}_{\nu'\nu}\int\tilde{G}_2[\gamma_{\nu\nu'}(\tau)]  \right) 
+\ldots\,. \label{eq:dabsrhodtrexpansion}
\end{align}
The complicated form of the operators $\tilde{G}_k$ prevents us from finding conditions for the validity of the first-order approximation.  We will establish conditions for the validity of the zeroth-order approximation.  

The $r$-dependent term is negligible relative to the $r$-independent term when 
\begin{align}
r & \ll \left|\frac{\mathrm{Re}\,w_0^{\nu\nu'}}{\mathrm{Re}\,w_1^{\nu\nu'}}\right| \notag \\
& = \left|
1+ (k_BT)^{1/2}
\frac{2^{3/2}m^{1/2}[\hbar(b_{\nu}^{\mathrm{i}}+b_{\nu'}^{\mathrm{i}})-2(\alpha_{\nu}\alpha_{\nu'}+\beta_{\nu}\beta_{\nu'})]}{\pi^{1/2}\hbar(\beta_{\nu}+\beta_{\nu'})} 
+ \right. \notag \\
& 
\left. \qquad \qquad \qquad+
k_BT\frac{6m[\hbar^2(c_{\nu}^{\mathrm{i}}+c_{\nu'}^{\mathrm{i}})+2\hbar\mathrm{Re}(a_{\nu}b_{\nu'}^{*}+b_{\nu}a_{\nu'}^{*})]}{\hbar^2(\beta_{\nu}+\beta_{\nu'})}+\ldots \right|
\,. \label{eq:firstsmallrcondition}
\end{align}
The coefficient of the $T^{1/2}$ dependent term in Eq.~\eqref{eq:firstsmallrcondition} is positive (except in the special case when both the numerator and denominator vanish, which we do not consider further), thus at low temperatures the mass ratio $r$ at which the zeroth-order approximation is valid increases with increasing temperature.  The $T^{1/2}$-dependent term is insignificant with respect to the $T$-independent term when
\begin{align}
(k_{B}T)^{1/2} & \ll \frac{\pi^{1/2}\hbar(\beta_{\nu}+\beta_{\nu'})}{2^{3/2}m^{1/2}[\hbar(b_{\nu}^{\mathrm{i}}+b_{\nu'}^{\mathrm{i}})-2(\alpha_{\nu}\alpha_{\nu'}+\beta_{\nu}\beta_{\nu'})]} \,. \label{eq:firstTconditionsmallr}
\end{align}
When Eq.~\eqref{eq:firstTconditionsmallr} holds then the condition for validity of the zeroth-order approximation, Eq.~\eqref{eq:firstsmallrcondition} becomes
\begin{align}
r & \ll 1\,.\label{eq:rll1}
\end{align}
The exponential decay constant for the zeroth-order approximation for the unscaled time variable $t$ is then
\begin{align}
\lambda_1 & = -\frac{\hbar n_{\mathrm{gas}}l}{m}\mathrm{Re}\,w_0^{\nu\nu'} \simeq n_{\mathrm{gas}} 
\frac{2\pi\hbar(\beta_{\nu}+\beta_{\nu'})}{m}\,,\label{eq:lambda1}
\end{align}
where the validity of Eq.~\eqref{eq:firstTconditionsmallr} has enabled us to drop the $T^{1/2}$-dependendent and higher terms in the small-$T$ expansion of $\lambda_1$. The factor $\hbar n_{\mathrm{gas}}l/m$ comes from the conversion from the $\tau$ time scale to $t$.  At zeroth-order in $r$, the decrease in $|\rho_{\nu\nu'}^{\mathrm{el}}|$ is determined entirely by the leading order terms of the inelastic scattering cross-sections.

It is possible that, as in Sec.~\ref{sec:tempdependence}, there is a cutoff time for the validity of the zeroth-order approximation because of the non-exponential forms of Eqs.~\eqref{eq:drhodtaurexpansion} and \eqref{eq:dabsrhodtrexpansion}.  However, because the complicated forms of the $\tilde{G}_k$ preclude solving for $\gamma_{\nu\nu'}(t)$, we cannot estimate the length of time for which the approximation is valid.

\subsection{Relative coherence in trapped sample}

Differentiating Eq.~\eqref{eq:etadef} and using Eq.~\eqref{eq:drhodtaurexpansion}, we obtain
\begin{align}
\frac{d}{d\tau}\eta_{\nu\nu'}(\tau) & = \omega_0 \eta_{\nu\nu'}(\tau) + r\omega_1 \eta_{\nu\nu'}(\tau) +\notag \\
&\;+ r^2\left(
\frac{\int\,\tilde{G}_2[\gamma_{\nu\nu'}(\tau)]}{\rho_{\nu\nu'}(\tau)}
+\frac{\int\,\tilde{G}_2[\gamma_{\nu'\nu}(\tau)]}{\rho_{\nu'\nu}(\tau)}
-\frac{\int\,\tilde{G}_2[\gamma_{\nu\nu}](\tau)}{\rho_{\nu\nu}(\tau)}
-\frac{\int\,\tilde{G}_2[\gamma_{\nu'\nu'}](\tau)}{\rho_{\nu'\nu'}(\tau)}
\right)\eta_{\nu\nu'}(\tau) +\ldots \label{eq:ddtetalowr}
\end{align}
where
\begin{align}
\omega_k 
&= \frac{w^{\nu\nu'}_k +w^{\nu'\nu}_k -w^{\nu\nu}_k -w^{\nu'\nu'}_k}{2} \notag \\
& = -\frac{\theta^{1/2}}{2\pi^{3/2}l^2}
\int d^3q\,q(4-2q^2)^ke^{-q^2}\int d^2\uvec{n}\,\left|f_{\nu\nu}\left(\frac{\hbar\theta^{1/2}}{l}q\uvec{n},\frac{\hbar\theta^{1/2}}{l}\gvec{q}\right)
-f_{\nu'\nu'}\left(\frac{\hbar\theta^{1/2}}{l}q\uvec{n},\frac{\hbar\theta^{1/2}}{l}\gvec{q}\right)
\right|^2 \notag \\
& \leq 0
\end{align}
for $k=0$ and 1.  As before, the decay of coherence is exponential up to first order in $r$ regardless of the temperature, and we cannot determine conditions for the validity of the first-order approximation.
We find
\begin{align}
\omega_0 & = -\frac{4\pi^{1/2}\theta^{1/2}}{l^2} |a_\nu-a_{\nu'}|^2      
+\frac{6\pi\hbar\theta}{l^3}\mathrm{Re}\left[(a_{\nu}-a_{\nu'})(b_{\nu}^*-b_{\nu'}^*)\right]
+ \ldots
 \\
\omega_1 & =  
-\frac{6\pi\hbar\theta}{l^3}
\mathrm{Re}\left[(a_{\nu}-a_{\nu'})(b_{\nu}^*-b_{\nu'}^*)\right] \notag \\
& \qquad \qquad \qquad
 +\frac{16\pi^{1/2}\theta^{3/2}\hbar^{2}}{l^4}
 \left\{|b_{\nu}-b_{\nu'}|^2 -2\mathrm{Re}\left(a_{\nu}-a_{\nu'})(c_{\nu}^*-c_{\nu'}^*)\right] \right\}
+\ldots
\,.
\end{align}

The $r$-dependent term is negligible compared to the $r$-independent term when
\begin{align}
r &\ll \left|\frac{\omega_0}{\omega_1}\right| \notag \\
& = 
\left| \frac{-2|a_\nu-a_{\nu'}|^2}{3\pi^{1/2}(2mk_BT)^{1/2}\mathrm{Re}\left[(a_{\nu}-a_{\nu'})(b_{\nu}^{*}-b_{\nu'}^{*})\right]} \right. \notag \\
& \qquad \left.
+1 - \frac{8
|a_\nu-a_{\nu'}|^2\left\{|b_{\nu}-b_{\nu'}|^2 -2\mathrm{Re}[(a_{\nu}-a_{\nu'})(c_{\nu}^{*}-c_{\nu'}^{*})]\right\}
}{9\pi
\left\{\mathrm{Re}\left[(a_{\nu}-a_{\nu'})(b_{\nu}^{*}-b_{\nu'}^{*})\right]\right\}^2
}+\ldots
\right|
\label{eq:rcondition}
\end{align}
The last expression is an expansion at small $T^{1/2}$ using Eq.~\eqref{eq:fexpansion}, which has been obtained with the ultracold temperature assumption that there is only $s$-wave scattering.  The terms after the first may be ignored when
\begin{align}
& (2mk_BT)^{1/2} \ll 
\frac{
\frac{2|a_\nu-a_{\nu'}|^2}{3\pi^{1/2}\left|\mathrm{Re}[(a_{\nu}-a_{\nu'})(b_{\nu}^{*}-b_{\nu'}^{*})]\right|}
}{\left|
1 - \frac{8
|a_\nu-a_{\nu'}|^2\left\{|b_{\nu}-b_{\nu'}|^2 -2\mathrm{Re}[(a_{\nu}-a_{\nu'})(c_{\nu}^{*}-c_{\nu'}^{*})]\right\}
}{9\pi \left\{\mathrm{Re}\left[(a_{\nu}-a_{\nu'})(b_{\nu}^{*}-b_{\nu'}^{*})\right]\right\}^2
}\right|}\,,\label{eq:secondTcondition}
\end{align}
in which case the condition Eq.~\eqref{eq:rcondition} for the validity of the zeroth-order in $r$ approximation becomes 
\begin{align}
m^{3/2} & \ll 
\frac{2|a_\nu-a_{\nu'}|^2M}{3\pi^{1/2}(2k_BT)^{1/2}|\mathrm{Re}[(a_{\nu}-a_{\nu'})(b_{\nu}^{*}-b_{\nu'}^{*})]|}\,.
\end{align} 
The exponential decay constant for $\eta_{\nu\nu}$ in the zeroth-order approximation for the unscaled time variable $t$ is
\begin{align}
\lambda_2 & = -\frac{\hbar n_{\mathrm{gas}}l}{m}\omega_0 \notag \\
& =
\frac{ n_{\mathrm{gas}}}{m}   \Bigl( 
\left(2mk_BT\right)^{1/2} 4\pi^{1/2} |a_\nu-a_{\nu'}|^2      
-(2mk_BT) 6\pi\mathrm{Re}\left[  (a_{\nu}-a_{\nu'})(b_{\nu}^{*}-b_{\nu'}^{*})  \right]
\Bigr. \notag \\
&\Bigl. \qquad \qquad \quad
+ (2mk_BT)^{3/2} 8\pi^{1/2}\left\{|b_{\nu}-b_{\nu'}|^2 -2 \mathrm{Re}[(a_{\nu}-a_{\nu'})(c_{\nu}^{*}-c_{\nu'}^{*})] \right\} + \ldots
\Bigr)
\,.\label{eq:lambda2}
\end{align}
Satisfying Eq.~\eqref{eq:secondTcondition} is not sufficient to determine the order of $(2mk_BT)^{1/2}$ at which this expansion may be terminated.  If the denominator on the right-hand side of Eq.~\eqref{eq:secondTcondition} is comparable to or greater than unity then all terms after the first may be ignored.

The remarks at the end of Sec.~\ref{sec:smallrabsrho} regarding possible cutoff times for the validity of the zeroth-order approximation also apply to the approximation in this section. 

\section{Conclusion}
  
The decoherence of trapped particles prepared in a coherent superposition of internal states as a result of collisions with buffer gas atoms at very low temperatures is described by Eqs.~\eqref{eq:ddtabsrho} and \eqref{eq:ddteta}.   The time evolution of the coherence between the internal states $\nu$ and $\nu'$ is parametrized by the complex scattering lengths of the particles in these two states and the other coefficients in the expansions of the $s$-wave scattering amplitudes $f_{\nu\nu}(p)$ and $f_{\nu'\nu'}(p)$ given in Eq.~\eqref{eq:fexpansion}.  Given these scattering parameters, Eqs.~\eqref{eq:ddtabsrho} and \eqref{eq:ddteta} can be used to calculate the decoherence rates taking into account elastic collisions as well as inelastic contributions.  These equations and their time-integrated forms, Eqs.~\eqref{eq:absrhoversust} and \eqref{eq:etaversust}, show that coherence between the internal states of the trapped tracer molecules, and its time evolution, can be represented in the limit of low temperatures as an expansion in powers of $T^{1/2}$.  Regardless of the temperature, in both Eqs.~\eqref{eq:ddtabsrho} and \eqref{eq:ddteta}, terms with higher-order $T^{1/2}$-dependence become significant given sufficient time.  

Eq.~\eqref{eq:ddtabsrho} describes the time evolution of the magnitude of the off-diagonal reduced matrix element $|\rho^{\mathrm{el}}_{\nu\nu'}|$ which characterizes the total coherent signal between the internal states $\nu$ and $\nu'$.  This quantity depends on the size of the remaining trapped population relative to the original number of trapped molecules at $t=0$.  Eq.~\eqref{eq:ddtabsrho} is a low-$T$ expansion of $\frac{d}{dt}|\rho^{\mathrm{el}}_{\nu\nu'}|$ in powers of $T^{1/2}$.  The leading term in the expansion of $\frac{d}{dt}|\rho^{\mathrm{el}}_{\nu\nu'}|$ in powers of $T^{1/2}$ is a temperature-independent term arising from trap loss caused by inelastic collisions.  It is determined by the complex parts of the scattering lengths of the states $\nu$ and $\nu'$, which are measures of the low-temperature inelastic scattering cross-section (cf. Eq.~\eqref{eq:sigmain}).  If $\beta_{\nu}=\beta_{\nu'}=0$, then the inelastic cross-sections vanish at the leading order in the collision momentum $p$ and the lowest order term in $\frac{d}{dt}|\rho^{\mathrm{el}}_{\nu\nu'}|$ varies as $T^{1/2}$.  The $T$-independent term varies as $m_*^{-1}$ and the $T^{1/2}$-independent term varies as $m_*^{1/2}$, hence loss of coherent signal is slower for heavier tracer molecules or buffer gas atoms.  This is because the mean velocity of a heavier particle is slower for a given temperature and hence the collision rate is lower. 

Eqs.~\eqref{eq:Tcondition} and \eqref{eq:cutofftime} give conditions on temperature and time for which the decay of $|\rho_{\nu\nu'}^{\mathrm{el}}|$ may be approximated by the temperature-independent term of Eq.~\eqref{eq:ddtabsrho}.  The regions of $(t,T^{1/2})$ space satisfying these conditions are indicated in Fig.~\ref{fig1}. 

Eq.~\eqref{eq:ddteta} gives the time-dependence of $\eta_{\nu\nu'}(t)$, which is the fraction of coherence between states $\nu$ and $\nu'$ in the trapped ensemble relative to that of a pure state.  The quantity $\eta_{\nu\nu'}$ depends only on the state of the trapped ensemble and not on the size of the trapped sample relative to the original population.  As a consequence, the expansion of $\frac{d}{dt}\eta_{\nu\nu'}$ in powers of $T^{1/2}$ has no $T$-independent term.  The leading order term has $T^{1/2}$-dependence.  It is determined by the square magnitude of the difference in complex scattering lengths between the states $\nu$ and $\nu'$, $|a_\nu-a_{\nu'}|^2=(\alpha_\nu-\alpha_{\nu'})^2+(\beta_\nu-\beta_{\nu'})^2$ and therefore depends on both elastic and inelastic scattering properties.

If both of the scattering lengths are equal in their real and imaginary parts, Eq.~\eqref{eq:ddteta} shows that the $T^{1/2}$-dependent term of $\frac{d}{dt}\eta_{\nu\nu'}$ vanishes.  In this case the leading term will depend on a higher power of $T^{1/2}$.  These higher order terms will in general be nonzero since they are governed by the coefficients of terms of order $p$ and higher (e.g. $b_\nu$, $b_{\nu'}$) in the expansion given by Eq.~\eqref{eq:fexpansion} (see Eq.~\eqref{eq:sigmael}).  The decay of $\eta_{\nu\nu'}$ may be approximated by the leading order, $T^{1/2}$-dependent term of Eq.~\eqref{eq:ddteta} when Eqs.~\eqref{eq:ineqcondition1} and \eqref{eq:ineqcondition2} hold.  These conditions are illustrated in Fig.~\ref{fig2}.  The $T^{1/2}$-dependent term varies as $m_*^{1/2}$.

For low bath particle/tracer particle mass ratios $m/M$, the decay of coherence measured by either $|\rho_{\nu\nu'}^{\mathrm{el}}|$ or $\eta_{\nu\nu'}$ is found to be exponential up to first order in $m/M$, regardless of temperature.  When Eqs.~\eqref{eq:firstTconditionsmallr} and \eqref{eq:rll1} are satisfied, the zeroth-order approximation for the evolution of $|\rho_{\nu\nu'}^{\mathrm{el}}|$ may be applied.  In this approximation decay is exponential with decay constant $\lambda_{1}$ given by Eq.~\eqref{eq:lambda1}.  The zeroth-order approximation, with exponential decay with decay constant $\lambda_{2}$ given by Eq.~\eqref{eq:lambda2}, may be applied to $\eta_{\nu\nu'}$ when Eqs.~\eqref{eq:secondTcondition} and \eqref{eq:rcondition} hold.     

The scattering lengths $\alpha_\nu$ for ultracold atoms are usually found from thermalization measurements \cite{monroe:prl,schmidt:prl,hopkins:pra}, which yield results with significant error bars.  Decoherence rates can be measured sensitively by observing the damping rate of coherent oscillations \cite{reinhold:nimprb,vrakking:pra,ramakrishna:prl}.  The imaginary parts of the scattering lengths $\beta_{\nu}$ are relatively easy to find precisely by trap loss measurements.   Since $\xi_1$ and $\lambda_2$ (at low temperatures) are proportional to $(\alpha_\nu-\alpha_{\nu'})^2+(\beta_\nu-\beta_{\nu'})^2$, Eqs.~\eqref{eq:ddteta} or Eqs.~\eqref{eq:ddtetalowr} and \eqref{eq:lambda2} could therefore provide the basis for a method of precise determination of the real part of the scattering length $\alpha_{\nu'}$ when the other scattering length $\alpha_\nu$ is known.  Precise measurements of scattering lengths in ultracold gases may be used as a probe of fundamental constants and symmetries in nature \cite{chin:njp}.

\begin{acknowledgments}
We thank J.\ Weinstein and B.\ Vacchini for helpful comments. The work was supported by NSERC of Canada.
\end{acknowledgments}

\appendix

\section{}

Here we evaluate the integral in Eq.~\eqref{eq:inteval22}.   From Eqs.~\eqref{eq:ic} and  \eqref{eq:G1def}, and using $\gamma^{\nu\nu'}_0(0)=\gamma_{\nu\nu'}(0)$, we obtain
\begin{align}
\int G_1[G_1[\gamma^{\nu\nu'}_0(0)]] & =  \left[\frac{2\pi i \hbar (b_\nu-b^*_{\nu'})+4\pi a_\nu a^*_{\nu'}}{l^2}\right]^2
\frac{A(r)}{r^{3/2}\pi^{9/2}}\rho_{\nu\nu'}(0) \label{eq:Aevaln}
\end{align}
where
\begin{align}
A(r) & = \int d^3Q\,e^{-\frac{Q^2}{r}}\left(\int d^3q\,qe^{-(\gvec{Q}+\gvec{q})^2}\right)^2\,. \label{eq:Adefn}
\end{align}
Integrating with respect to $\gvec{Q}$ yields
\begin{align}
A(r) & = \frac{r^{3/2}\pi^{3/2}(2r+1)^{5/2}}{(r+1)^4}A_1\left(\frac{r}{r+1}\right) \label{eq:Aeqn}
\end{align}
where 
\begin{align}
A_1(s) & = \int d^3x\,d^3y\,xy \exp\left[-\left(x^2+y^2-2s\gvec{x}\cdot\gvec{y}\right)\right] \notag \\
& = \frac{\pi^2}{s}\frac{d^2}{ds^2}A_2(s)\label{eq:diffs}
\end{align}
with
\begin{align}
A_2(s) & = \int^\infty_0 dx\int^\infty_0 dy\, e^{-(x^2+y^2)}\left(e^{2sxy}-e^{-2sxy}\right)\,.
\end{align}
We convert to polar coordinates $(R,\theta)$ and integrate with respect to $R$. We then make the change of variables $u=s(1-s^2)^{-1/2}\cos2\theta$ and integrate with respect to $u$, obtaining $A_2(s)=(1-s^2)^{-1/2}\sin^{-1}s$.  Performing the differentiations of Eq.~\eqref{eq:diffs}, we substitute the result into Eq.~\eqref{eq:Aeqn}, followed by substitution into Eq.~\eqref{eq:Aevaln} to obtain Eq.~\eqref{eq:inteval22}.

\end{document}